\documentclass[11pt,a4paper]{article}
\usepackage{jcappub}
\usepackage{graphicx}
\usepackage{amsmath}
\usepackage{epstopdf}
\usepackage{latexsym}
\usepackage{dcolumn}
\usepackage{bm}

\input{epsf}
\newcommand{\be}{\begin{equation}}
\newcommand{\ee}{\end{equation}}
\newcommand{\ba}{\begin{eqnarray}}
\newcommand{\ea}{\end{eqnarray}}
\newcommand{\bi}{\begin{itemize}}
\newcommand{\ei}{\end{itemize}}
\newcommand{\ga}{\gtrsim}
\newcommand{\bfi}{\begin{figure}
\epsfxsize=9cm
\epsffile}
\newcommand{\bfinew}{\begin{figure}
\begin{center}
\includegraphics}
\newcommand{\efi}{\end{figure}}
\newcommand{\efinew}{
\end{center}
\end{figure}}
\newcommand{\no}{\nonumber}
\newcommand{\la}{\lesssim}

\newcommand{\muk}{\mu {\rm K}}

\title{Testing eternal inflation with the kinetic Sunyaev Zel'dovich effect} 
\author[a,b,c]{Pengjie Zhang}
\author[d,e]{and Matthew C. Johnson}
\affiliation[a]{Center for Astronomy and Astrophysics, Department of
  Physics and Astronomy, Shanghai Jiao Tong University, 955 Jianchuan road, Shanghai, 200240, China}
\affiliation[b]{IFSA Collaborative Innovation Center, Shanghai Jiao Tong
University, Shanghai 200240, China}
\affiliation[c]{Key Laboratory for Research in Galaxies and Cosmology, Shanghai Astronomical Observatory, 80 Nandan Road, Shanghai, 200030, China}
\affiliation[d]{Department of Physics and Astronomy, York University \\ Toronto, On, M3J 1P3, Canada}
\affiliation[e]{Perimeter Institute for Theoretical Physics \\ Waterloo, Ontario N2L 2Y5, Canada}

\emailAdd{zhangpj@sjtu.edu.cn}
\emailAdd{mjohnson@perimeterinstitute.ca}

\abstract{Perhaps the most controversial idea in modern cosmology is that our observable universe is contained within one bubble among many, all inhabiting the eternally inflating multiverse. One of the few way to test this idea is to look for evidence of the relic inhomogeneities left by the collisions between other bubbles and our own. Such relic inhomogeneities will induce a coherent bulk flow over Gpc scales. Therefore, bubble collisions leave unique imprints in the cosmic microwave background (CMB) through the kinetic Sunyaev Zel'dovich (kSZ) effect, temperature anisotropies induced by the scattering of photons from coherently moving free electrons in the diffuse intergalactic medium. The kSZ signature produced by bubble collisions has a unique directional dependence and is tightly correlated with the galaxy distribution; it can therefore be distinguished from other contributions to the CMB anisotropies. An important advantage of the kSZ signature is that it peaks on arcminute angular scales, where the limiting factors in making a detection are instrumental noise and foreground subtraction. This is in contrast to the collision signature in the primary CMB, which peaks on angular scales much larger than one degree, and whose detection is therefore limited by cosmic variance. In this paper, we examine the prospects for probing the inhomogeneities left by bubble collisions using the kSZ effect. We provide a forecast for detection using cross-correlations between CMB and galaxy surveys, finding that the detectability using the kSZ effect can be competitive with constraints from CMB temperature and polarization data. }

\begin{document}
\maketitle

\section{Introduction}

The idea that we might inhabit an eternally inflating multiverse is surely one of the most radical concepts to come out of modern physics. Although radical, accelerated expansion of the universe and spontaneous symmetry breaking, both observed phenomena in nature, are the necessary and sufficient ingredients. Once accelerated expansion driven by vacuum energy begins, an epoch known as inflation in the early universe, it is generically difficult to end globally. Instead, inflation ends only locally inside of bubbles which are formed only rarely via quantum mechanical effects. If bubbles are formed slower than the space between them is expanding, inflation cannot end globally, giving rise to the phenomenon of eternal inflation.   

In this picture, our universe is housed within one bubble universe among many, perhaps each with different properties and cosmological histories (as motivated by the string theory landscape~\cite{Susskind:2003kw}). The Coleman de Luccia instanton~\cite{Coleman:1977py,Callan:1977pt,Coleman:1980aw} that mediates bubble nucleation dictates the symmetry of  a one-bubble spacetime, giving rise to an open FRW universe in the bubble interior. If we live inside a bubble, a second epoch of slow-roll inflation is necessary to dilute curvature to observable levels and provide the seeds for structure formation~\cite{Gott:1982zf,Gott:1984ps,Bucher:1994gb}.  

Each bubble undergoes collisions with others~\cite{Garriga:2006hw}, producing potentially observable wreckage~\cite{Aguirre:2007an}. Assessing the character of this wreckage and the probability of observing it has been the subject of a substantial body of work~\cite{Hawking:1981fz,Hawking:1982ga,Wu:1984eda,Moss:1994pi,Aguirre:2007wm,Aguirre:2008wy,Aguirre:2009ug,Kozaczuk:2012sx,Chang_Kleban_Levi:2009,Chang:2007eq,Czech:2010rg,Freivogel_etal:2009it,Gobbetti_Kleban:2012,Kleban_Levi_Sigurdson:2011,Kleban:2011pg,Wainwright:2013lea,Johnson:2010bn,Johnson:2011wt,Freivogel:2007fx,Wainwright:2014pta,Salem:2012gm,Czech:2011aa,Salem:2011qz,Salem:2010mi,Larjo:2009mt,Easther:2009ft,Giblin:2010bd,Ahlqvist:2014uha,Kim:2014ara,Ahlqvist:2013whn,Hwang:2012pj,Deskins:2012tj,Amin:2013dqa,Amin:2013eqa}. Given an underlying model consisting of a scalar field theory coupled to gravity, a full set of predictions can be made for the primordial metric fluctuations caused by a bubble collision~\cite{Wainwright:2013lea,Wainwright:2014pta}. The symmetry of the collision spacetime forbids the production of gravitational waves~\cite{Hawking:1982ga,Wu:1984eda}, and therefore the effects of a collision in any single-field model can be encoded in the primordial scalar comoving curvature perturbation. Bubble collisions can be mapped onto specific inhomogeneous initial conditions for slow-roll inflation inside of the bubble. Because the wreckage produced by bubble collisions is pre-inflationary, having the minimum number of inflationary $e-$folds is necessary for observable effects to be accessible today~\footnote{Various arguments may be advanced for the plausibility of this situation, see e.g.~\cite{Freivogel:2005vv,Yang:2012jf}}. 

Each collision is an event, and therefore can only affect regions in its causal future. The bubble interior is split into regions that are affected by the collision and regions that are not. Observers in the vicinity of this causal boundary have access to the most distinctive observational signatures; in the following we will focus on this class of observers. The first constraints on cosmic bubble collisions in the eternal inflation scenario were found by Feeney et al~\cite{Feeney_etal:2010dd,Feeney_etal:2010jj,Feeney:2012hj,McEwen:2012uk} using observations of the Cosmic Microwave Background (CMB) radiation made by the WMAP satellite~\cite{Bennett:2003ba}. This was followed up by the analysis of Osbourne et al~\cite{Osborne:2013hea,Osborne:2013jea}. Neither of these analyses support the bubble collision hypothesis, placing a constraint on the possible models of eternal inflation. However, these analyses were performed with an incomplete understanding of the collision signature and the connection between the underlying theory and observable predictions. An analysis incorporating the improved predictions of~\cite{Wainwright:2013lea,Wainwright:2014pta} is underway.

\bfinew[width=9cm]{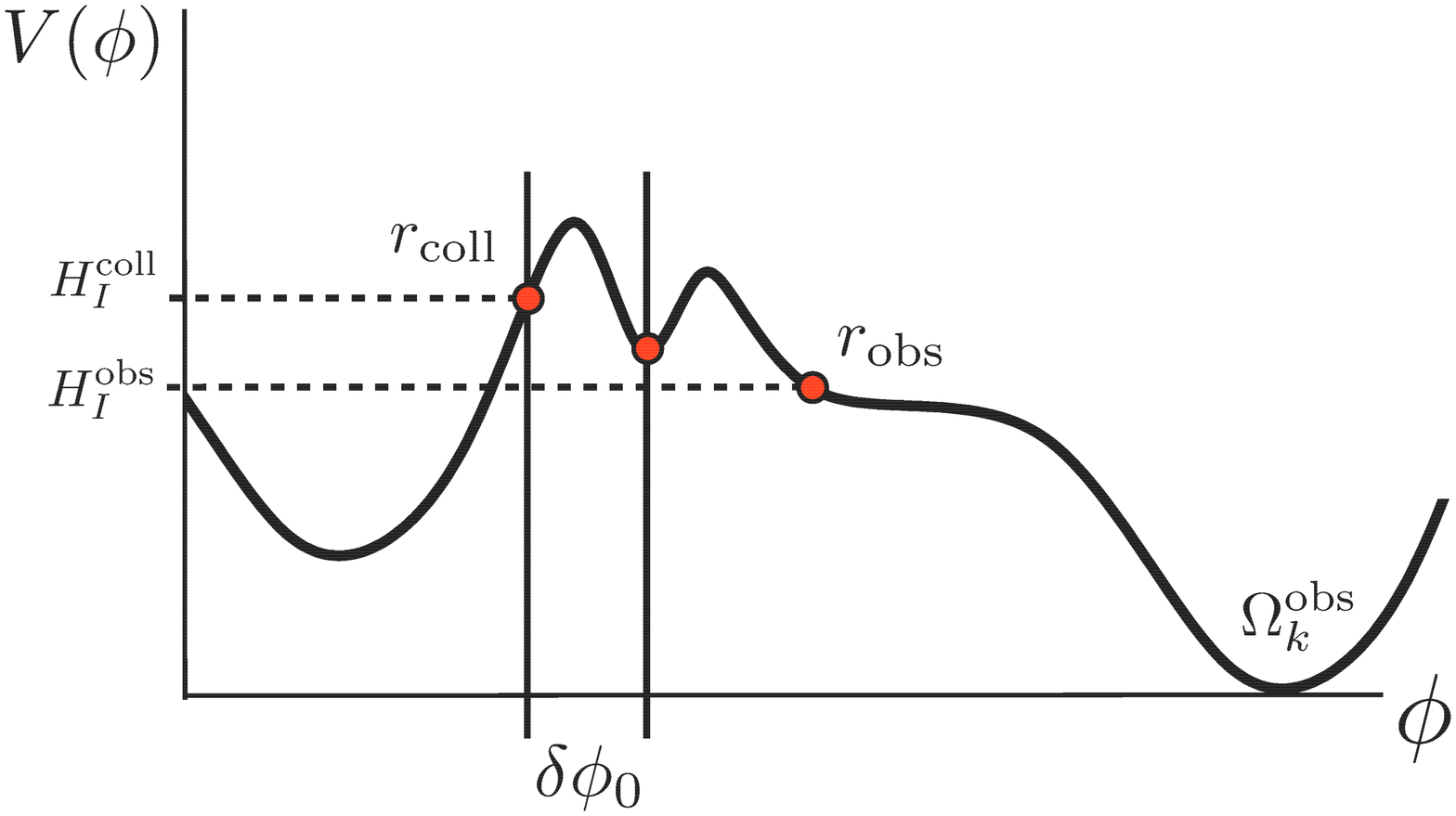}
\caption{The scalar field potential underlying the bubble collision model. \label{fig:potential}}
\efinew
\bfinew[width=9cm]{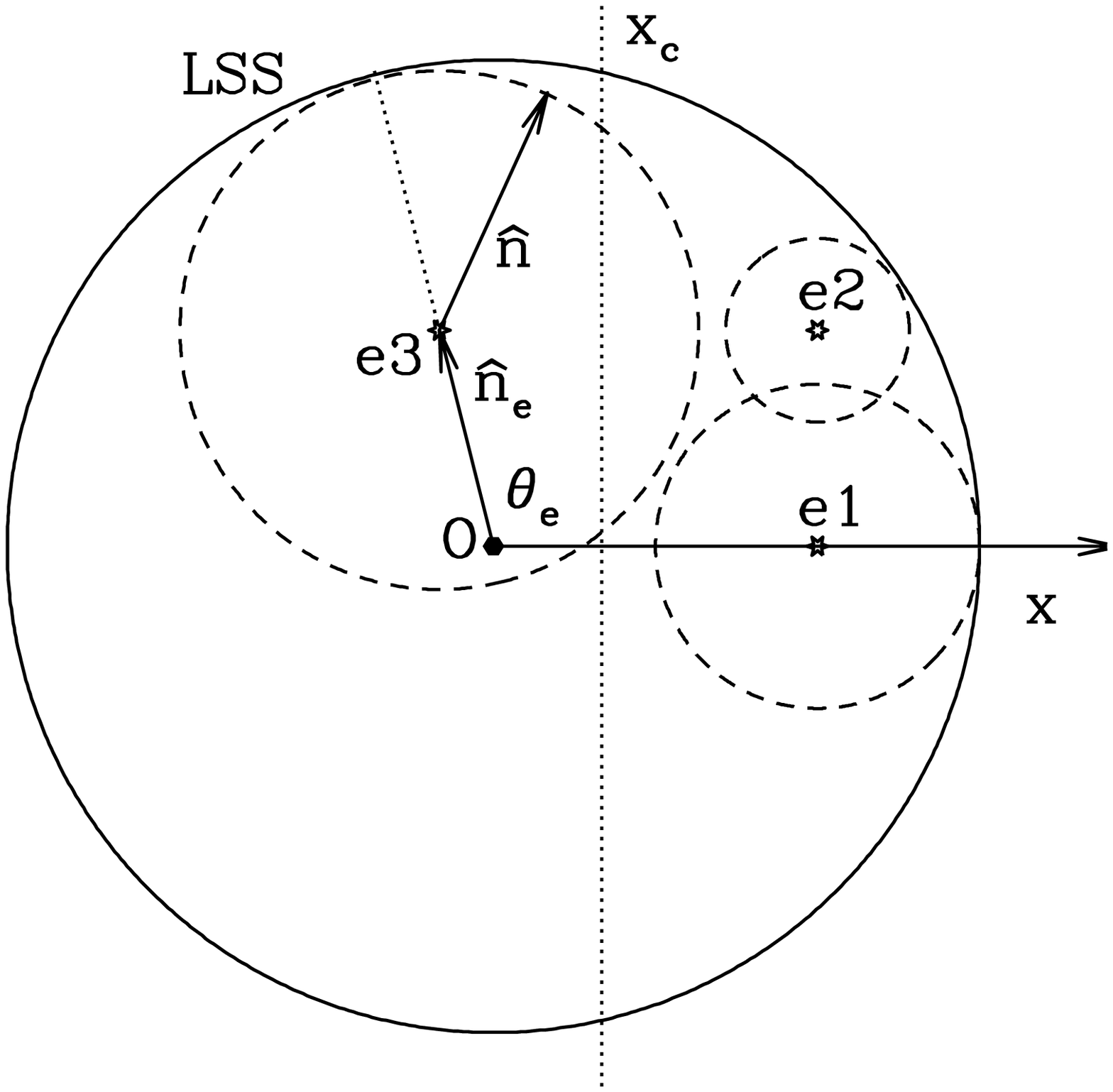}
\caption{A schematic diagram of the kSZ effect in the collision spacetime. The causal boundary of the collision $x_c$ is denoted by the dotted line (plane). The solid circle (sphere) is
  the last scattering surface (LSS) seen by us, and the dashed circles (spheres) are the LSS of several representative free electrons, labelled by  ``e1'', ``e2'' and ``e3'' respectively. These electrons see 
  corresponding CMB dipoles on their LSS and have their own peculiar
  motion with respect to our CMB. In contrast to a single adiabatic superhorizon mode,
  the bubble collision generated perturbation induces $v_{\rm eff}\neq
  0$ and therefore induces a non-vanishing kSZ effect. Unlike the primary CMB
  fluctuation generated by bubble collision, which cover only a
  fraction of the sky, the kSZ effect generated by the
  bubble collision is spread over the whole $4\pi$ sky. For example, all electrons 
at a radius between ``e3'' and us (``O'') at all viewing angles can see the bubble
  collision through their LSS and hence generate a kSZ effect visible
  to us.   \label{fig:bc}}
\efinew

Because the dominant contribution to the CMB temperature anisotropies arises from the time of last scattering, it is not an ideal probe of inhomogeneity. CMB polarization probes the local temperature quadrupole observed by electrons, providing  important additional information on inhomogeneities (see e.g.~\cite{Dvorkin:2007jp,Mortonson:2009qv}) and acting as an important discriminator between anisotropy and inhomogeneity. However, since the signature of a bubble collision is imprinted on the largest scales, cosmic variance is a significant impediment to placing stringent constraints on the bubble collision model. Can we ever hope to do better? 

This paper explores the ability of another probe of large-scale inhomogeneities, the kinetic Sunyaev Zel'dovich (kSZ) effect \cite{SZ80}, to test the bubble collision hypothesis. The kSZ effect is caused by the inverse Compton scattering of CMB photons by coherently moving free electrons. Its power to probe missing baryons, reionization and peculiar velocity at $\la 100$ Mpc scale has been long recognized (e.g. \cite{Vishniac87,Haehnelt96,Ma02,Zhang04a,Zhang08b}), and has been explored in CMB experiments such as ACT and SPT (e.g. \cite{Hand12,George14}). It can also be utilized to constrain dark matter-dark energy interaction \cite{Xu13} and modified gravity \cite{Xu14,Mueller14}. Recently, a new regime of the kSZ effect generated by  larger scale ($\ga$ Gpc)  inhomogeneities has been realized and applied to  probe dark flows~\cite{Kashlinsky08,Zhang10d,Kashlinsky11,Li12,Lavaux13,Atrio-Barandela14,Planck-I-2014-v}, test the Copernican principle \cite{Goodman95,Zhang11b}, and constrain the present day vacuum decay rate~\cite{Pen14}. This paper presents its newest  application. 

The kSZ effect arises from scattering of CMB photons from free-electrons undergoing bulk motion in the late universe. The amplitude of the effect is proportional to the locally observed CMB dipole, and is therefore a sensitive measure of bulk velocities.\footnote{Although we do not consider it in this paper, polarization of the kSZ photons map the locally observed CMB quadrupole~\cite{Kamionkowski:1997na}.} Horizon scale inhomogeneities generate bulk motions coherent over Gpc scales with characteristic directional dependence, and therefore, a distinctive kSZ signal. Note that this is a special class of kSZ signatures, distinctive from the conventional kSZ effect used to probe peculiar velocities on small scales. 

The kSZ effect is a unique and powerful probe of horizon scale inhomogeneities, for a number of reasons.  
\begin{enumerate}
\item Since the distance to the epoch of reionization (at $z\sim 6-10$) is very close to the distance to our horizon, free electrons probe much of the observable universe. Therefore the kSZ effect provides a unique opportunity to observe the universe from virtually every position inside of our horizon, and is sensitive to horizon scale inhomogeneities such as those generated by a bubble collision.
\item The kSZ signature is essentially a modulated re-mapping of the electron distribution to the CMB. Crucially, due to small scale inhomogeneities in the electron distribution, the power of the kSZ effect generated by horizon scale inhomogeneities peaks at arcminute angular scales.\footnote{Roughly speaking, fluctuation in such kSZ effect is proportional to the matter density fluctuation projected  along some radial depth. Therefore the peak scale depends on the linear transfer function, $\sigma_8$, relevant redshifts, and the projection length. For typical LCDM density fluctuation projected along Gpc radial depth, the typical peak scale is of the order one arcminute. This is numerically demonstrated in cases of dark flow induced kSZ effect \cite{Zhang10d}) and the kSZ effect discussed in this paper (Fig. \ref{fig:cl}). }  Hence, the kSZ effect promises to give constraints beyond the cosmic variance limited results obtained from the primary CMB only.
\item In contrast to the conventional kSZ effect, whose two-point cross correlation with the galaxy density field vanishes, the new kSZ effect is tightly correlated with the galaxy density field \cite{Zhang10d,Zhang11b}. Furthermore, the overall amplitude of the cross correlation has a unique directional modulation, depending on specific models of horizon scale inhomogeneities. Hence by specially designed cross correlation with galaxies, known as kSZ tomography \cite{Shao11b},  we can not only isolate it from other CMB components (primary CMB, other  kSZ components, cosmic infrared background, etc), but also determine its redshift dependence.
\end{enumerate}

The remainder of the paper is structured as follows. In Sec.~\ref{sec:bubblecollisions} we discuss the predicted inhomogeneities produced by bubble collisions. Sec.~\ref{sec:ksz_for_bubbles} computes the kSZ signature expected for a general bubble collision.  Sec.~\ref{sec:tomography} estimates the capability of probing bubble collision using kSZ tomography, and makes a comparisong with constraints obtained from the primary CMB. Finally, we discuss and summarize in Sec.~\ref{sec:discussion}. 

\section{Cosmology in the aftermath of a bubble collision}\label{sec:bubblecollisions}

A model of eternal inflation is defined by a scalar field lagrangian, which by assumption has a metastable false vacuum and at least one true vacuum. Transitions from the false to the true vacuum are governed by the CDL instanton, which gives rise to bubbles of true vacuum nucleating in the false vacuum. The cosmological history inside the bubbles includes an epoch of cosmic inflation, and the O(4) symmetry of the CDL instanton guarantees that a single bubble contains an infinite open Friedman Robertson Walker (FRW) universe. Focusing on a single bubble, which we hypothesize might contain our observable universe, additional bubbles forming from the false vacuum will collide with ours forming a fractal cluster of intersecting bubbles~\cite{Guth:1982pn,Garriga:2006hw}. In the limit where collisions are rare, they will lie outside of each other's horizon; we can therefore treat collisions on a pair-wise basis. The spacetime describing the collision between two bubbles has an SO(2,1) symmetry~\cite{Wu:1984eda}, inherited from the symmetry of each of the colliding bubbles. Determining the precise outcome of a collision, and therefore the observable effects, has been examined using analytic and numerical methods. Recent efforts by Wainwright et al~\cite{Wainwright:2013lea,Wainwright:2014pta} have established a connection between the primordial curvature perturbation and specific lagrangians, forming a first-principles derivation of the effects of bubble collisions in single field models. The conclusions of this work are as follows:
\begin{itemize}
\item The boundary of the causal future of the collision splits the inner-bubble cosmology into two qualitatively different regions. This work considered observers who are in the vicinity of this boundary, reconstructing the perturbed FRW universe in this region in comoving gauge.
\item On the scale of our observable universe, where spatial curvature can be neglected, the collision spacetime has approximate planar symmetry.
\item Neglecting spatial curvature (at most a $1\%$ effect), the perturbed FRW universe in Newtonian gauge is
\begin{equation}
ds^2=-(1+2\Psi)dt^2+a^2(t)(1-2\Psi)d{\bf x}^2
\end{equation}
with
\be
\label{eqn:iPhi}
\Psi_i({\bf r})\simeq \left\{
\begin{array}{cl}
A(\tilde{x}-\tilde{x}_c)+B(\tilde{x}-\tilde{x}_c)^2, &
{\rm if}\ x\geq x_c\\
0,& {\rm if  }\ x<x_c\ .
\end{array}\right.
\ee
Here, we have fixed the collision direction to be along the $x$ axis, and $x_c$ is the location of the causal boundary. The prior probability distribution over $x_c$ is flat over the size of the present day horizon. $\tilde{x}\equiv x/r_H$ and $r_H\equiv c  H_0^{-1}$ is the present day Hubble radius (where $c$ is the speed of light). 
\item A comparison between numerical and analytic methods in~\cite{Wainwright:2014pta} showed that the coefficients $A$ and $B$ can be related to fundamental parameters of the scalar field lagrangian shown in Fig.~\ref{fig:potential} by~\footnote{Comparing with Ref.~\cite{Wainwright:2014pta}, a factor of $3/5$ has been applied to convert between the Newtonian and comoving gauges, and we have converted our dimensionless measure of distance from the distance to the surface of last scattering to the size of the present day Hubble radius, introducing another factor of 3.}:
\begin{eqnarray}\label{eq:AandB}
A &=&  \frac{2}{5} \sqrt{ \frac{8 \Omega_k^{\rm obs}}{r_{\rm obs}} } \frac{\delta \phi_0^{\rm coll}}{M_{\rm pl}} \left( 1 - \cos \Delta x_{\rm sep} \right) \\
B &=& \frac{2}{15} \Omega_k^{\rm obs} \sqrt{ \frac{r_{\rm coll}}{r_{\rm obs}} }   \frac{H_{I}^{\rm coll}}{H_{I}^{\rm obs}} \left( 1 - \cos \Delta x_{\rm sep} \right)^2 \nonumber
\end{eqnarray}
where $\delta \phi_0$ is the distance between the instanton endpoints connecting the false vacuum to the collision bubble interior, $\Omega_k^{\rm obs}$ is the energy density in curvature in our own universe (constrained to be $\Omega_k^{\rm obs} \lesssim 10^{-2}$), $r_{\rm obs}$ is the tensor to scalar ratio for inflation inside the observation bubble (constrained to be $r_{\rm obs} \lesssim .1$), $r_{\rm coll}$ is the tensor to scalar ratio for inflation inside the collision bubble, $H_{I}^{\rm obs}$ is the Hubble scale during inflation inside the observation bubble, $H_{I}^{\rm coll}$ is the Hubble scale during inflation inside the collision bubble, and $\Delta x_{\rm sep}$ is the initial proper distance between the bubbles in the collision frame measured in terms of the false vacuum Hubble scale ($0 < \Delta x_{\rm sep}< \pi$). By definition, $\delta \phi_0$ is positive for collisions between identical bubbles and negative for collisions between different bubbles. The prior probability distribution over initial separations $\Delta x_{\rm sep}$ is proportional to $\sin^3 \Delta x_{\rm sep}$. We therefore expect $(1 - \cos \Delta x_{\rm sep}) \sim \mathcal{O}(1)$. The analytic expression Eq.~\ref{eq:AandB} assumes that slow-roll inflation is not disrupted by the collision, in which case $H_{I}^{\rm obs}$ is roughly constant everywhere in the collision spacetime during inflation. Complete disruption of inflation will occur when $\delta \phi_0^{\rm coll}$ is comparable to the total field excursion during inflation inside the bubble. Using the Lyth Bound~\cite{Lyth:1996im} ($\sqrt{r_{\rm obs}} = \sqrt{0.01} \Delta \phi / M_{\rm pl}$), we therefore conclude that Eq.~\ref{eq:AandB} will be valid as long as $\delta \phi_0^{\rm coll} < M_{\rm pl} \sqrt{r_{\rm obs} / 0.01}$.
\item We restrict ourselves to so-called native born observers~\cite{Aguirre:2009ug}, that is those who are comoving with respect to the unperturbed portion of the bubble. By definition, this restricts $x_c > 0$. There are also observers comoving with respect to the perturbed portion of the bubble. In the vicinity of $x_c$, these observers would have causal access to the collision boundary. Here, and also deep within the collision region, the dominant perturbation would be a quadratic curvature perturbation centred on the observer. 
\end{itemize}

Cosmological datasets can be used to constrain the parameters in the template Eq.~\ref{eqn:iPhi}, which in turn constrains the underlying scalar field lagrangian through Eq.~\ref{eq:AandB}. We now turn to assessing the ability of the kSZ signature produced by bubble collisions to be used as such a probe.

\section{The kinetic Sunyaev Zel'dovich effect generated by bubble
  collision}\label{sec:ksz_for_bubbles}
The curvature perturbation induced by bubble collisions gives rise to relative motion between matter in different regions of the Universe. Inverse Compton scattering off free electrons in relative motion induces temperature anisotropies in the observed CMB sky via the kSZ effect. The relevant induced temperature anisotropies are given by~\cite{SZ80}
\be
\label{eqn:kSZ}
\frac{\Delta T}{T_{\rm CMB}}(\hat{n}_e)=- \sigma_T \int_0^{\chi_{\rm re}} \ d\chi_e \ a_e \ 
n_e(\chi(z_e)\hat{n},z_e) \ \frac{v_{\rm
    eff}(\hat{n}_e)}{c}   \ .
\ee 
The minus sign is chosen such that when an electron moves away from us
($v>0$), the kSZ effect is negative. The vector $\hat{n}_e$ points from
us to the CMB sky, which is also the vector from us to the free
electrons scattering CMB photons. The subscript ``e'' denotes properties of free electrons. $a_e=1/(1+z_e)$ is the scale factor. $\chi_e$ is the comoving radial coordinate of free electrons, and $d\chi_e = c dz_e/H$. For a flat universe, $\chi_e$ coincides with the comoving angular diameter distance. $n_e$ is the electron number density. $v_{\rm eff}$ is the effective
velocity generating the kSZ effect, which we derive in detail below. Besides the peculiar velocity of electrons at redshift $z$ and direction $\hat{n}_e$, it also takes into account the contribution from the intrinsic CMB dipole on the last scattering surface (LSS) of free electrons at different locations. This new contribution only exists in non-Copernican universes.  

A schematic plot of the kSZ effect produced in the bubble collision spacetime is shown in Fig. \ref{fig:bc}. Since the distance to the reionization epoch at $z\sim 10$ is very close to the distance to our horizon, free electrons are essentially everywhere inside of our horizon. Each electron sees its own LSS, and combining all these LSS enables us to fairly sample all regions inside of our horizon.  For example, the primary CMB along the line of sight from ``O'' to ``e3'' is not affected by the depicted bubble collision. Nevertheless, the kSZ effect generated by electrons located between ``e3'' and ``O'' allows us to infer the existence of a bubble collision along this line of sight. At fixed distance from us, the kSZ effect is confined within a disk. However, combining electrons from all distances spreads the effects of the bubble collision over the whole sky. 

\subsection{The effective velocity responsible for the kSZ effect}
To calculate $v_{\rm eff}$, we need to determine the evolution of the Newtonian potential and velocity. The perturbation induced by bubble collisions (Eq.~\ref{eqn:iPhi}) is a mixture of both superhorizon and subhorizon modes, with a dominant contribution from superhorizon modes. Subhorizon modes are further damped compared to superhorizon modes during cosmological evolution, with the transfer function satisfying $T(k\gg 1/H)/T(k\ll 1/H)\ll 1$. Therefore we expect that the overall evolution of the potential and peculiar velocity can be approximated by that of the superhorizon modes, which are scale independent, and have analytical expressions (e.g. \cite{Erickcek08}). We outline this computation in Appendix~\ref{sec:appendix}. To test this approximation, we numerically compute the potential and peculiar velocity by solving the Boltzmann equations for radiation and cold dark matter (CDM) in a flat universe.  We confirm that the superhorizon approximation works very well (Fig. \ref{fig:v}, Appendix~\ref{sec:appendix}). Including the sub-horizon evolution acts to slightly smooth the causal boundary at $x_c$, a consequence of the damping of subhorizon (small scale) modes. We neglect this effect, as well as the baryon-photon coupling before last scattering~\footnote{This gives rise to the ``cosmic wakes" described in Ref.~\cite{Kleban_Levi_Sigurdson:2011}, and may add structure to baryon peculiar velocities on BAO scales in the vicinity of the causal boundary beyond what is considered here. We do not expect such signatures to significantly alter our results.}. Hereafter we will adopt the superhorizon approximation and use the analytic expressions in Appendix~\ref{sec:appendix} to calculate the kSZ effect induced by bubble collision. 

The potential $\Psi$ at late times follows the same form as Eq. \ref{eqn:iPhi} but with a scale-invariant growth factor $D_\Psi (a)$ (given by Eq. \ref{eqn:PhiSH}):
\begin{equation}
\Psi({\bf r},a) = D_{\Psi} (a) \Psi_i({\bf r})\ .
\end{equation}
For the velocity field, only the x-component is non-zero. It is well described by 
\be
\label{eqn:vdis}
v_x({\bf r},a) = - D_v(a)r^{-1}_H\times \left\{
\begin{array}{cl}
A+2B(\tilde{x}-\tilde{x}_c), &
{\rm if}\ x\geq x_c\\
0,& {\rm if  }\ x<x_c\ .
\end{array}\right.
\ee
The linear velocity growth factor $D_v$ is given by Eq. \ref{eqn:Dv}. 
Throughout the paper we will discuss two limiting cases with
$(A,B)=(\neq 0,0)$ (case A) and $(A,B)=(0,\neq 0)$ (case B).

\bfinew[width=9cm]{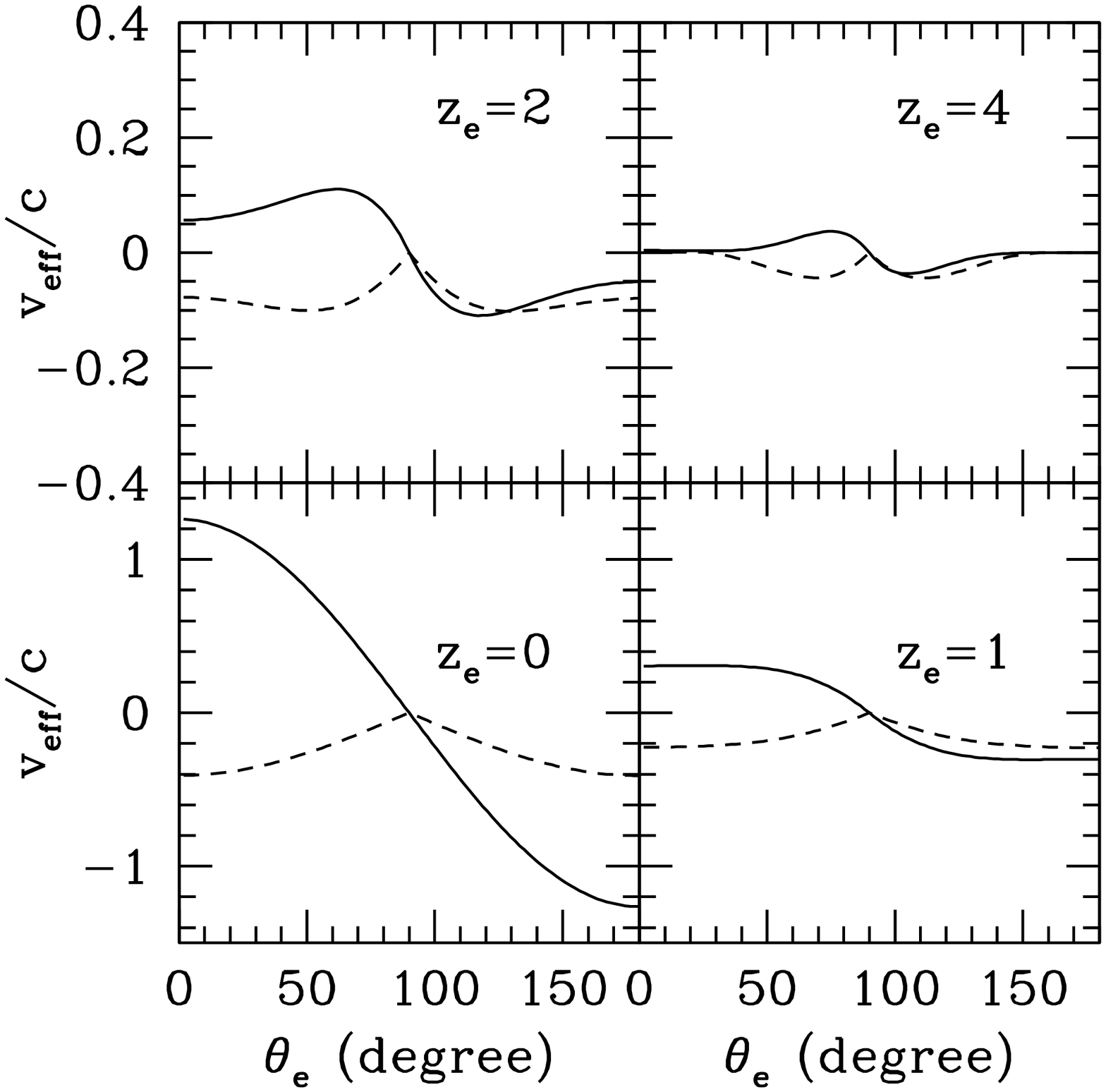}
\caption{The effective peculiar velocity that generates the kSZ
  effect, for $z_c=0$. Electrons sit at $z_e=0,1,2,4$ respectively. The solid lines have $B=1$ and $A=0$. The dash lines have $A=1$
  and $B=0$. $z_c$ is the closest redshift to the causal boundary of the collision. Due to the azimuthal symmetry of the initial 
  perturbation, $v_{\rm eff}=0$ when $\theta_e=\pi/2$. For $A=1$ and
  $B=0$,  we have the analytical expression at $z_e=0$: $v_{\rm eff}=v_e\cos(\theta_e)$ when $\theta_e>\pi/2$ and
  $-v_e\cos(\theta_e)$ when $\theta_e<\pi/2$. The value of $v_e$ is
  given in Fig. \ref{fig:v}. For $z_e=4$, $v_{\rm eff}$ approaches
  zero when $\theta_e\rightarrow 0$ or $\theta_e\rightarrow \pi$, but
  for different reasons. The reason it approaches zero when
  $\theta_e\rightarrow \pi$ is that the electrons and their last
  scattering spheres are located in unperturbed regions. The reason it
  approaches zero when $\theta_e\rightarrow 0$ is due to the
  cancellation of SW, ISW and Doppler components~\cite{Erickcek08}. \label{fig:zc0}}
\efinew
\bfinew[width=9cm]{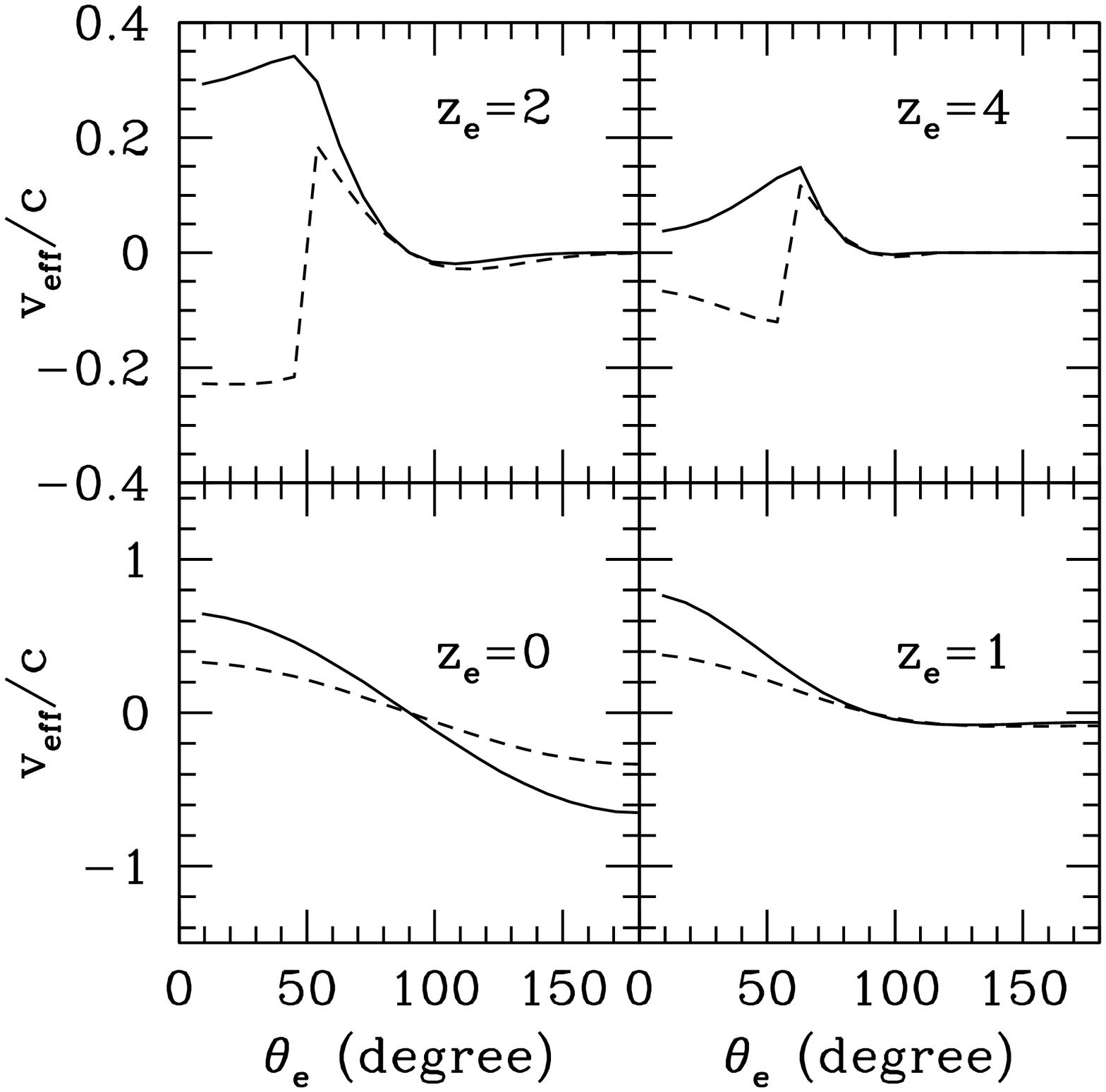}
\caption{Same as Fig. \ref{fig:zc0}, but for $z_c=1$. Notice that
  when $\theta_e\rightarrow \pi$ and $z_e\ga 2$, the electrons and their last
  scattering spheres are located in unperturbed regions and hence $v_{\rm
    eff}\rightarrow 0$.  The sharp decrease at $\theta_e\simeq
  50^{\circ}$ for $z_e=2$ (and at $\theta_e\simeq 60^{\circ}$ for
  $z_e=4$) and $B=0$ is caused by the discontinuity of velocity at the
  edge of bubble collision. This sharp feature shows up when $z_e>z_c$. $v_{\rm eff}(\hat{n})$ shows a rich dependence on the observation angle, a precious property for clean extraction from an otherwise 
  noisy CMB sky. \label{fig:zc1}}
\efinew

\bfinew[width=9cm]{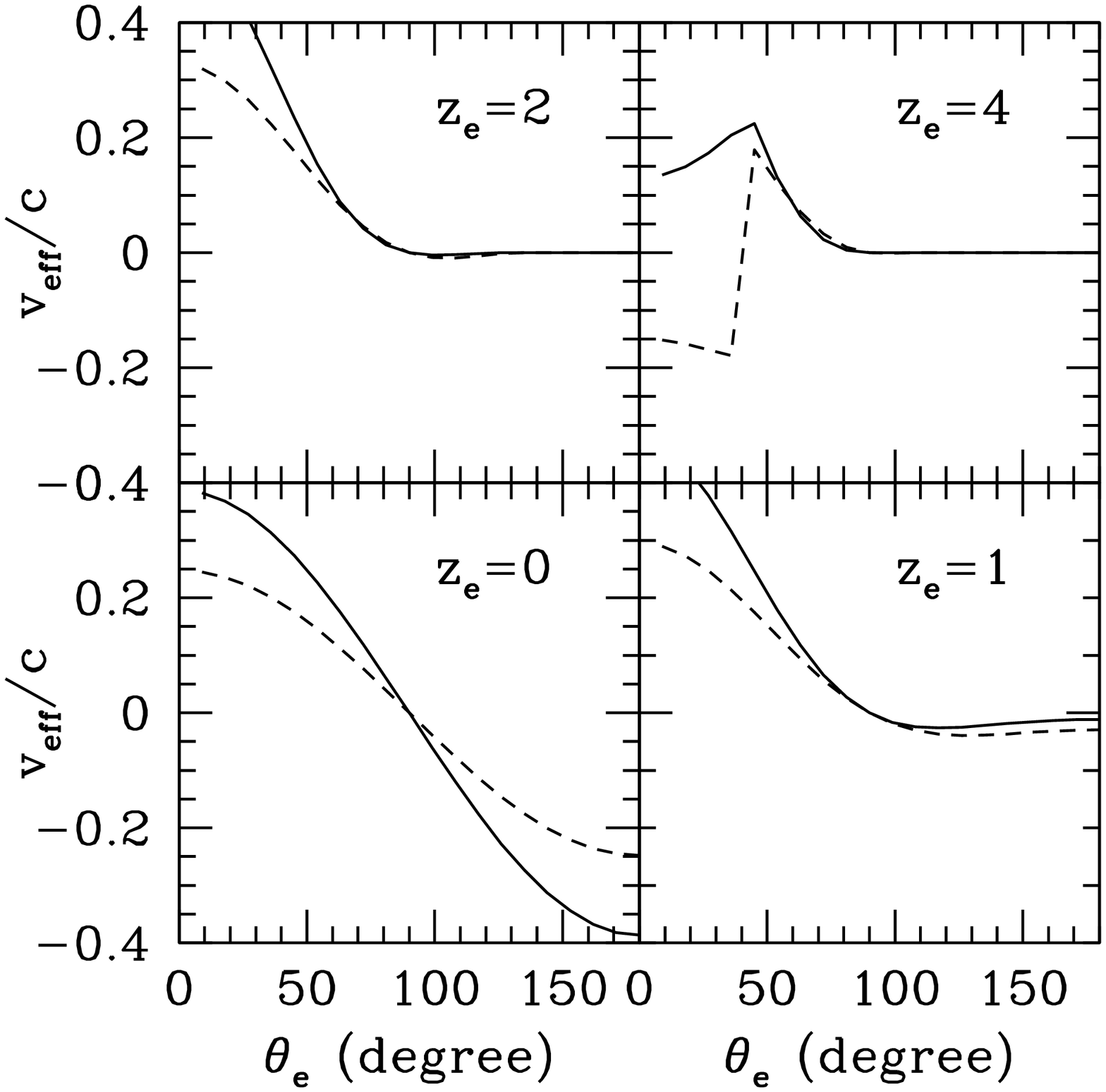}
\caption{Same as Fig. \ref{fig:zc0}, but for $z_c=2$. Notice that,
  for $z_e=0$, $v_{\rm eff}$ is anti-symmetric under the operation
  $\theta_e\leftrightarrow \pi-\theta_e$, a direct result of azimuthal
  symmetry. The only exception is when $z_c=0$ and $B=0$ where the anti-symmetry
  is reversed to symmetry, caused by the sharp transition in the
  velocity field at $z_e=z_c=0$ (Fig. \ref{fig:zc0}). \label{fig:zc2}}
\efinew
\bfinew[width=9cm]{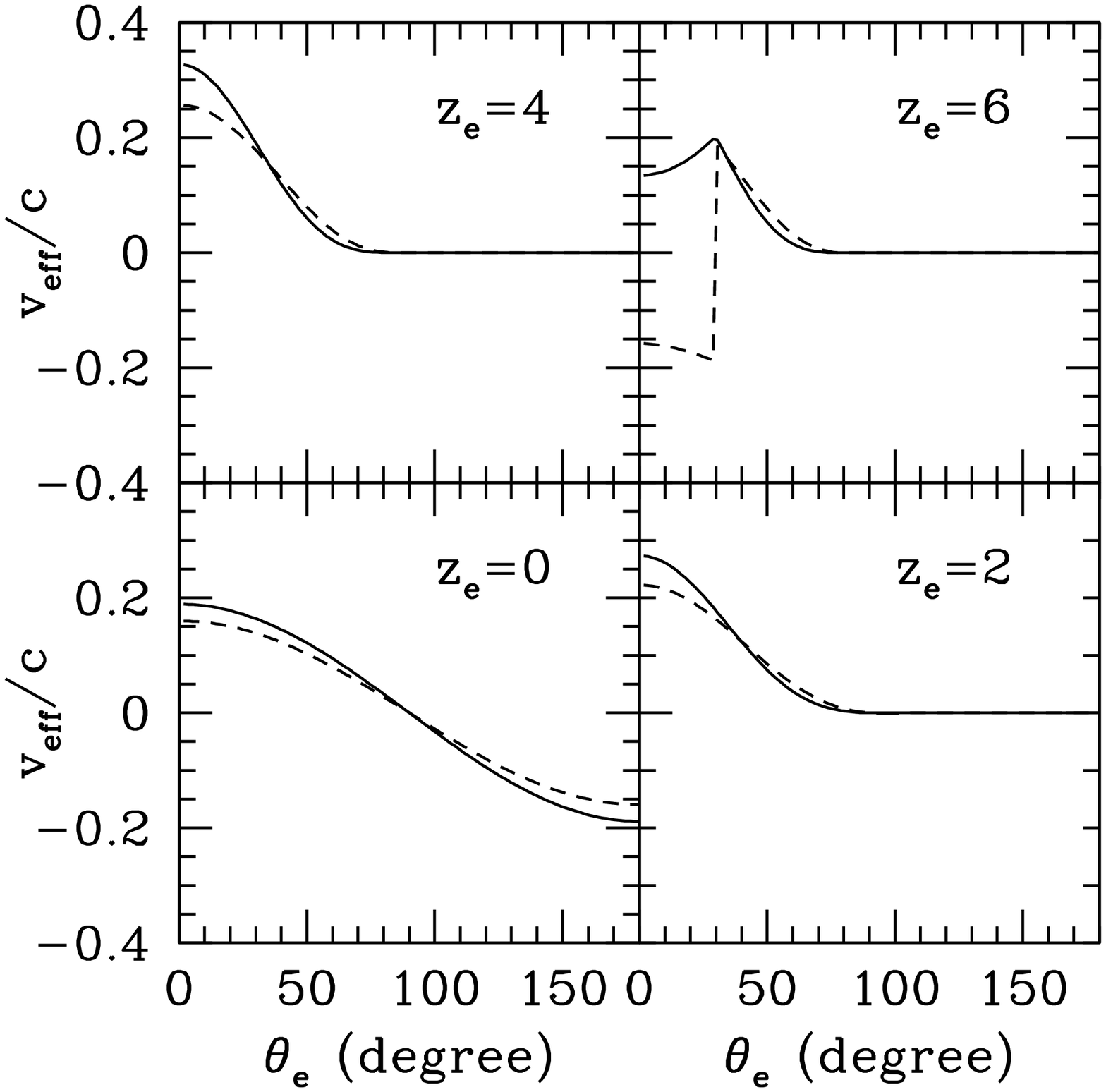}
\caption{Simiar to  Fig. \ref{fig:zc0}, but for $z_c=4$ and
  $z_e=0,2,4,6$ instead. The two cases
  of initial fluctuations produce similar $v_{\rm eff}$ at $z_e\la
  z_c$, but this is a coincidence. At higher $z_e$, the two cases differ
  significantly. In particular, the sharp transition at $z_e=6$ and
  $B=0$ shown in Fig. \ref{fig:zc2} shows up here when $z_e=6$. These electrons sit at the edge of bubble collision, satisfying $\chi_e\cos\theta_e=x_c\equiv \chi(z_c)$.   \label{fig:zc4}}
\efinew

\bfinew[width=9cm]{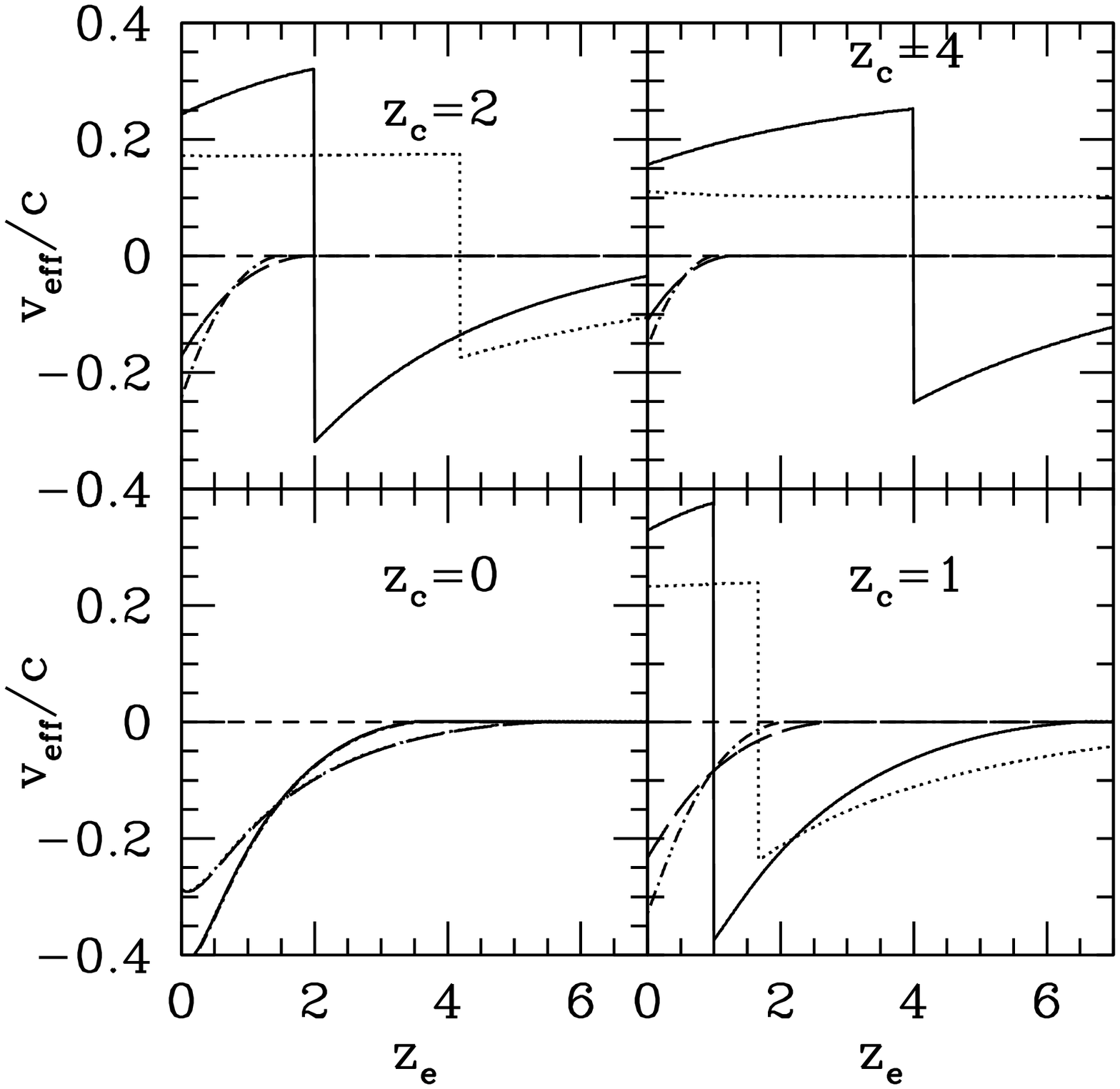}
\caption{The $v_{\rm eff}$-$z_e$ relation for $A=1$, $B=0$ and electrons with
  $\theta_e=0$ (solid lines), $\pi/4$ (dot lines), $\pi/2$ (dash lines), $3\pi/4$ (long dash
  lines), and $\pi$ (dot-short dash lines), respectively. Notice that due
  to the azimuthal symmetry, $v_{\rm eff}(\theta_e=\pi/2)=0$. For
  $z_c=0$, the lines with $\theta_e=0$ overlap with those for
  $\theta_e=\pi$. This can be understood by the result that $v_{\rm eff}=0$ if $\Psi=A(x-x_c)$ without the causal boundary at $x=x_c$ \cite{Erickcek08}. Also for this reason, the lines $\theta_e=\pi/4$ overlap with those for 
  $\theta_e=3\pi/4$. An interesting behavior is that for $\theta_e=0$, $v_{\rm
    eff}(z_e)$ is nearly a constant  when $z_e<z_c$. \label{fig:vza}}
\efinew

\bfinew[width=9cm]{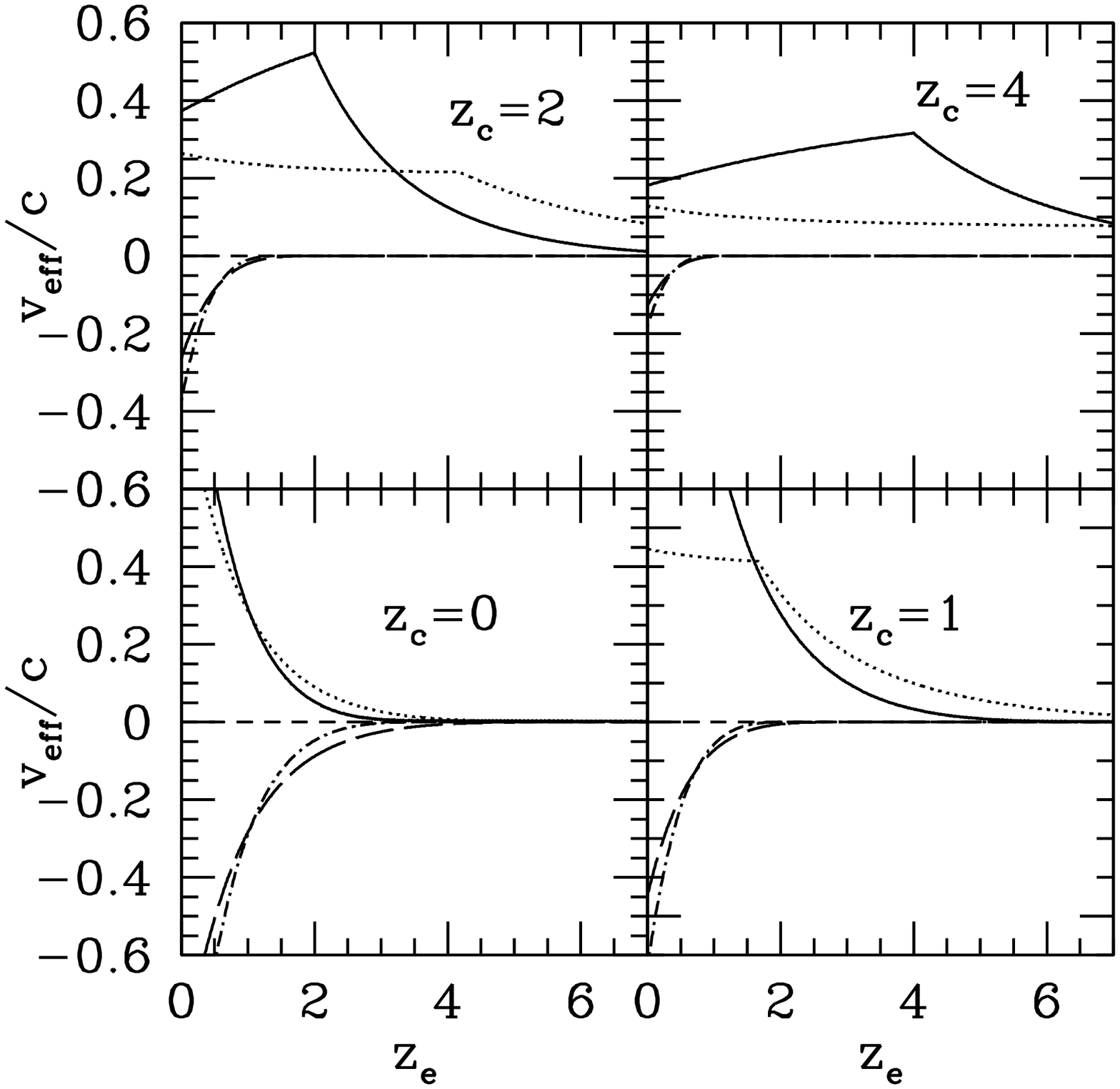}
\caption{Similar to Fig. \ref{fig:vza}, but for the $v_{\rm
    eff}$-$z_e$ relation with $B=1$ and $A=0$. \label{fig:vzb}} 
\efinew

\bfinew[width=9cm]{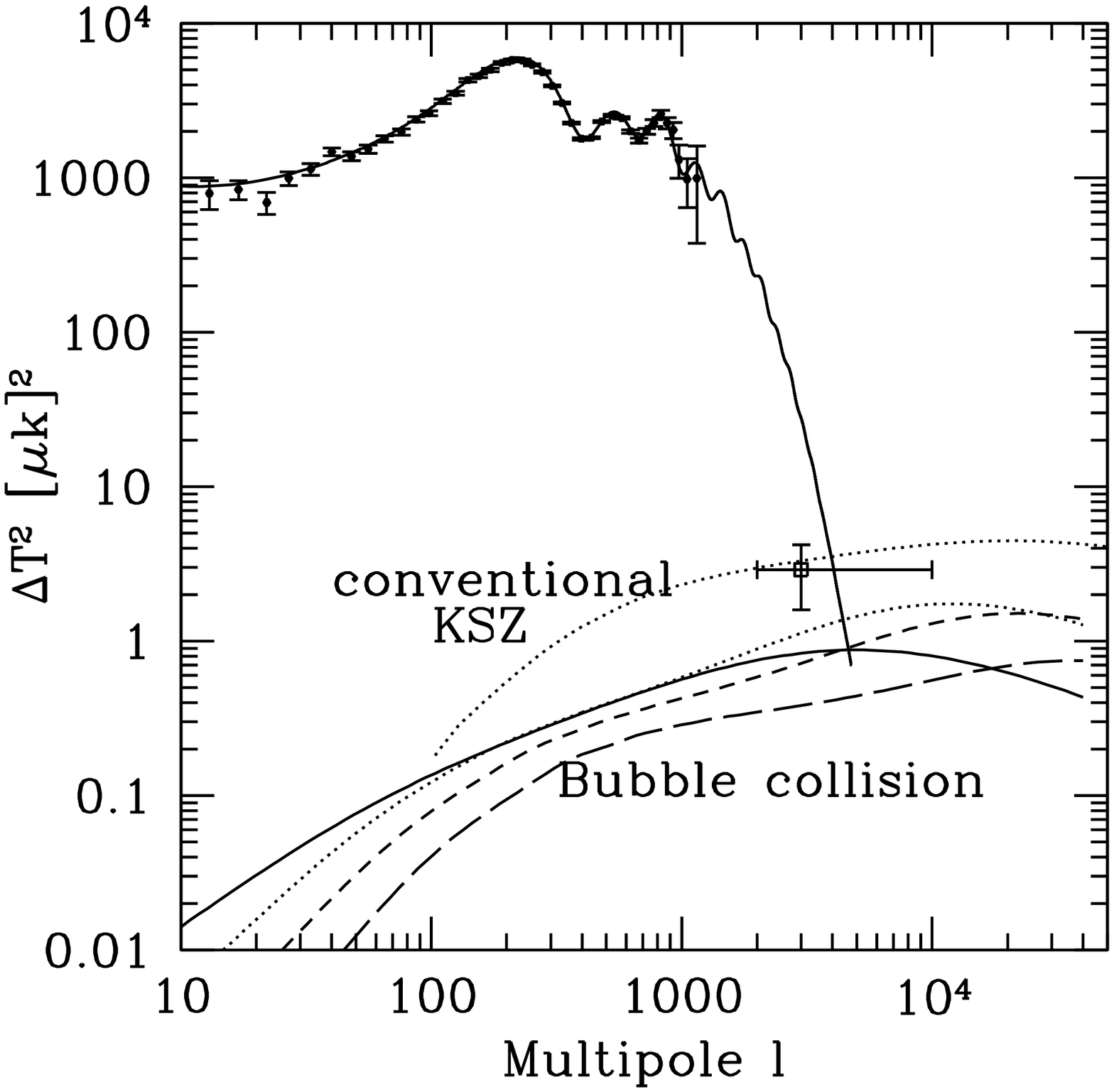}
\caption{The kSZ power spectrum for
  $B=10^{-3}$ along the line of sight towards the centre of the bubble collision. Solid, dot, shot dash and long dash lines have $z_c=0,1,2,4$ respectively. The generated kSZ effect is
  sub-dominant to the conventional kSZ effect calculated using the model of \cite{Zhang04a}. It is also well below the measured band-power of the conventional kSZ effect at $\ell\sim 3000$ ($2000<\ell<10000$) by
  SPT \cite{George14}. Given the existence of overwhelming contamination from a
  combination of the primary CMB, conventional kSZ and cosmic infrared
  background, it is infeasible to detect bubble collision through a measurement of the kSZ auto power spectrum. 
  \label{fig:cl}}
\efinew
\bfinew[width=9cm]{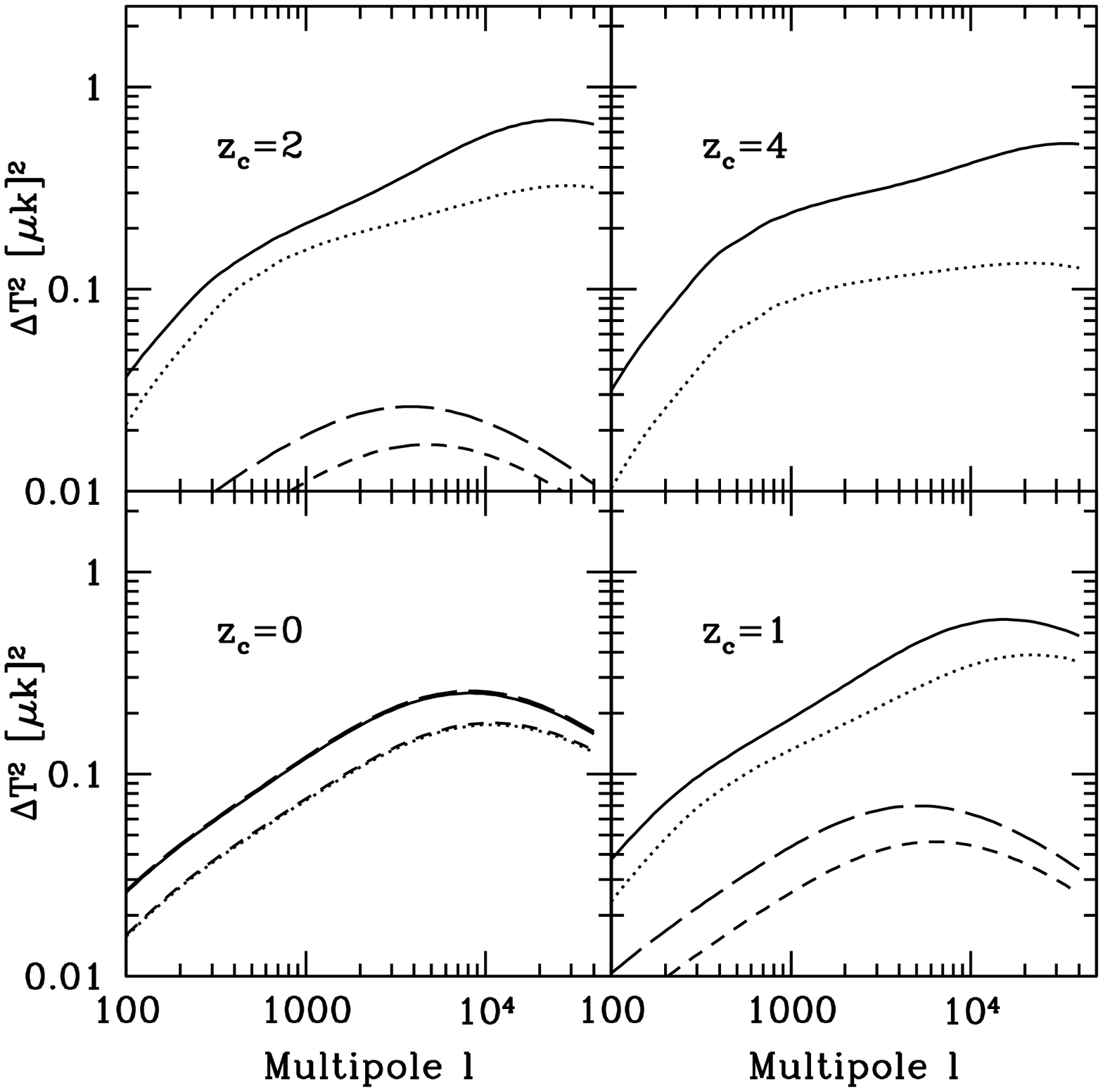}
\caption{The kSZ power spectrum for $A=10^{-3}$. Solid, dot, shot
  dash and long dash lines have $\theta_e=0^{\circ}, 45^{\circ},
  135^{\circ}, 180^{\circ}$ respectively. Notice that when
  $\theta_e=90^{\circ}$, the kSZ effect vanishes.  \label{fig:kszA}}
\efinew
\bfinew[width=9cm]{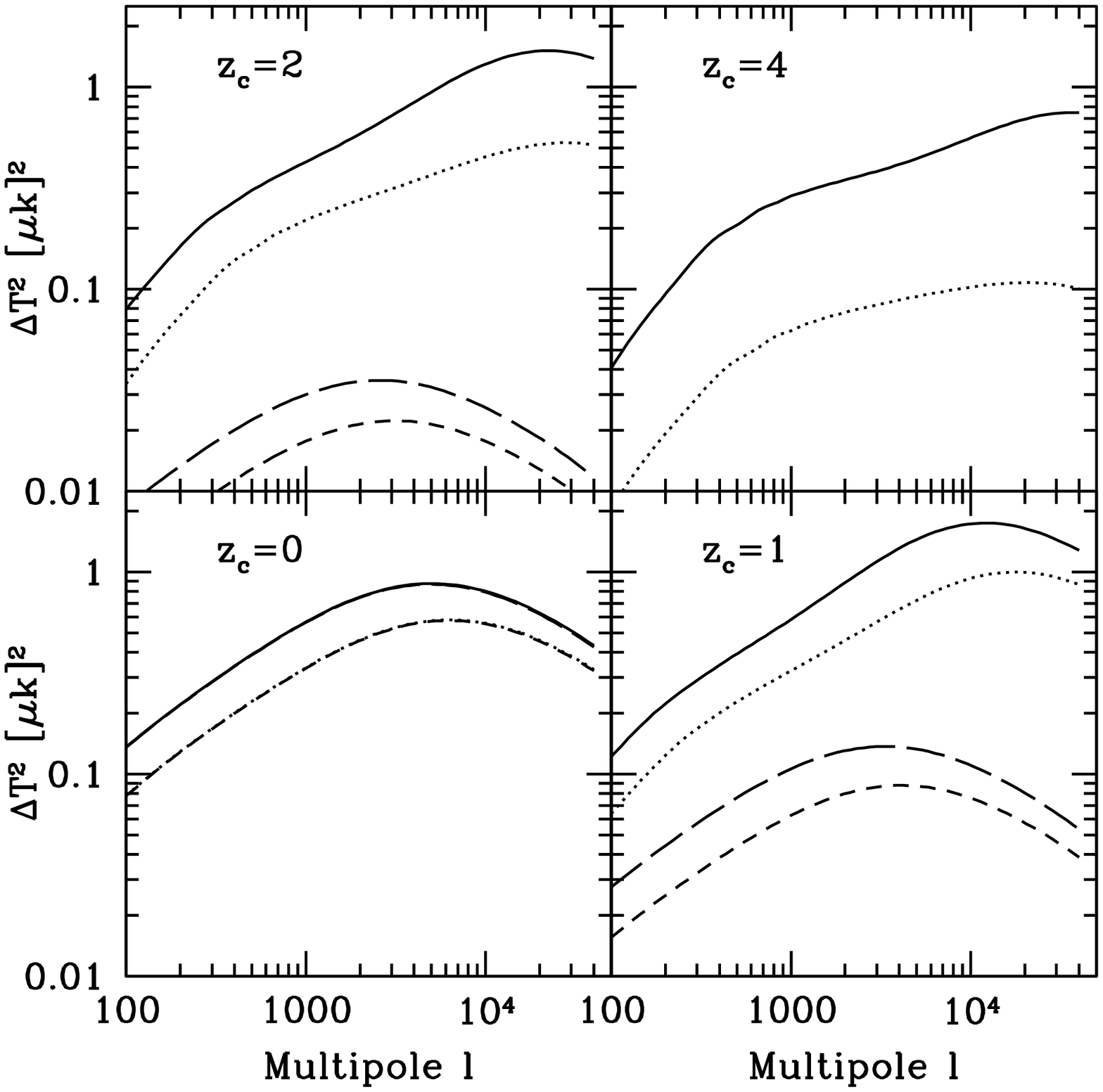}
\caption{The kSZ power spectrum for $B=10^{-3}$.  Solid, dot, shot
  dash and long dash lines have $\theta_e=0^{\circ}, 45^{\circ},
  135^{\circ}, 180^{\circ}$ respectively. Notice that when
  $\theta_e=90^{\circ}$, the kSZ effect vanishes. \label{fig:kszB}}
\efinew

The effective velocity in Eq.~\ref{eqn:kSZ} is related to the CMB dipole observed by a freely
falling electron via
\be\label{eq:veffdef}
v_{\rm eff}(\hat{n}_e)=\frac{3}{4\pi}\int_{4\pi} d^2\hat{n} \ \Theta({\bf r}_{\rm
    dec}) \ (\hat{n}\cdot\hat{n}_e) \ .
\ee
Here $\cos\theta\equiv \hat{n}\cdot\hat{n}_e$ and the unit vector
$\hat{n}$ points from the electrons to their LSS. ${\bf
  r}_{\rm dec}=\chi_e\hat{n}_e+\Delta \chi_{\rm
  dec}\hat{n}$. Here  $\chi_e\equiv \chi(z_e)$ is the distance from us
to free electrons, and $\Delta \chi_{\rm
  dec}\equiv \chi(z_{\rm dec})-\chi_e$ is the distance from the
electrons to their LSS. The geometry is shown in Fig.~\ref{fig:bc}. The effective velocity has three contributions, 
\be
v_{\rm eff}(\hat{n}_e)=v_{\rm eff,SW}+v_{\rm eff,Doppler}+v_{\rm
  eff,ISW}\ .
\ee
These contributions come from the Sachs-Wolfe (SW) effect generated by the
gravitational potential on the LSS, the
Doppler effect due to peculiar motion of photons on the LSS and peculiar
motion of electrons at redshift $z_e$, and the integrated Sachs-Wolfe (ISW)
effect \cite{SW67}, respectively. For a freely falling
electron at redshift $z_e$ and position ${\bf
  r}_e\equiv \chi_e\hat{n}_e$, the CMB temperature it
sees along the direction $\hat{n}$ is
\be
\Theta(\hat{n})=\Theta _{\rm SW} (\hat{n})+ \Theta_{\rm
  Doppler} (\hat{n})+\Theta _{\rm ISW}(\hat{n})\ .
\ee
We follow \cite{Erickcek08} to evaluate the above terms. The SW term is
\ba
\Theta_{\rm SW} (\hat{n})=\left[2-\frac{5}{3}\frac{9/10}{D_\Phi(a_{\rm dec})}\right]\Psi({\bf r}_{\rm dec},a_{\rm
  dec})=\left[2-\frac{5}{3}\frac{9/10}{D_\Phi(a_{\rm dec})}\right]D_\Psi(a_{\rm
  dec})\Psi_i({\bf r}_{\rm dec}) \ .
\ea
Here we have neglected the impact of $\Psi({\bf r}_e,a_e)$ since it
does not cause an observable effect.  The Doppler effect has two
 contributions with opposite sign, 
\ba
\Theta_{\rm
  Doppler} (\hat{n})=\hat{n}\cdot
[{\bf v}({\bf r}_e,a_e)-{\bf v}({\bf r}_{\rm dec},a_{\rm dec})]= \hat{n}\cdot \left[D_v(a_{\rm dec})\nabla \Psi_i({\bf r}_{\rm
    dec})-D_v(a_e)\nabla \Psi_i({\bf r}_e)\right]\ .
\ea
The ISW effect is given by 
\ba
\label{eqn:ISW}
\Theta_{\rm ISW} (\hat{n})=2\int_{a_{\rm
    dec}}^{a_e} \frac{d\Psi}{da}({\bf r}(a),a) da
=2\int_{a_{\rm
    dec}}^{a_e} \frac{dD_\Psi}{da}\Psi_i({\bf r}(a)) da \ .
\ea
Here, ${\bf r}(a)={\bf
  r}_e+\Delta \chi\hat{n}$. $\Delta \chi(a)\equiv \int_{a}^{a_e}
da/a^2H$. So $\Delta \chi_{\rm dec}=\Delta \chi(a_{\rm dec})$. 

Before proceeding, there is an important issue
to clarify. For adiabatic perturbations in a flat universe, and in the
limit of constant gradient in the 
gravitational potential  corresponding to a very 
superhorizon mode, the three contributions (SW, Doppler and ISW) cancel exactly ($\Theta(\hat{n})=0$) \cite{GZ78,Turner91,Bruni94,Erickcek08}. Therefore $v_{\rm eff}=0$ (vanishing CMB dipole). This was first found by
\cite{GZ78,Turner91,Bruni94} for a $\Omega_m=1$ Einstein-de Sitter
universe. Erickcek et al. \cite{Erickcek08} proved that it also holds
in an $\Lambda$CDM universe with radiation and an CDM universe with smooth dark
energy and radiation. They further argue that this should be a generic
feature for adiabatic superhorizon modes so that the cancellation holds
in all flat cosmologies.   For these cases,  the kSZ effect
vanishes. Fortunately for the bubble collision generated
inhomogeneities,  the exact cancellation no longer works, for two reasons. First, the
curvature perturbation has a quadratic term, and hence a spatially
varying gradient. So in general $v_{\rm
  eff}\neq 0$. Second, even if the quadratic term vanishes
($B=0$), due to the discontinuity in gradient on the edge of a bubble
collision, $v_{\rm eff}\neq 0$ for  electrons whose LSSs are not
completely inside of the region perturbed by  bubble collision (e.g. at position
``e3'' in Fig.~\ref{fig:bc}). Although electrons whose LSS are completely inside of the region affected by bubble collision (e.g.  ``e1'' and ``e2'' in Fig.~\ref{fig:bc}) can have $v_{\rm eff}=0$ for $B=0$, overall $v_{\rm
  eff}\neq 0$ and we expect a non-vanishing kSZ effect.

The azimuthal symmetry of the collision spacetime allows us to choose the parameterization
\begin{eqnarray}
\hat{n}_e &\equiv& (\cos\theta_e,\sin\theta_e,0)\ , \\
\hat{n} &\equiv&
(\cos\theta, \sin\theta\cos\varphi,\sin\theta\sin\varphi)\ . \no
\end{eqnarray}
 The integral over $\varphi$  in Eq.~\ref{eq:veffdef} is always $[0,2\pi]$.  To evaluate the integral over $\theta$ in Eq.~\ref{eq:veffdef}, we must consider three cases:
 \begin{enumerate}
 \item $\chi_e\cos\theta_e+\Delta  \chi_{\rm dec}<x_c$. The LSS of the electron is outside of the region affected by the bubble collision and therefore $v_{\rm eff}=0$. 
\item $\chi_e\cos\theta_e-\Delta  \chi_{\rm dec}>x_c$. The LSS of the electron is completely inside of the region affected by the bubble collision. So the integral is over $\theta\in[0,\pi]$.
\item$\chi_e\cos\theta_e-\Delta  \chi_{\rm 
  dec}<x_c$ \& $\chi_e\cos\theta_e+\Delta  \chi_{\rm
  dec}>x_c$. Only part of the LSS is inside of the region affected by the bubble collision. Therefore the integral is over $[0,\theta_c]$.  $\theta_c$ is the
solution of $\theta$ to $\chi_e\cos\theta_e+\Delta  \chi_{\rm
  dec}\cos\theta=x_c$, namely $\cos\theta_c=(x_c-\chi_e
\cos\theta_e)/\Delta \chi_{\rm dec}$. 
\end{enumerate}
 
For cases 2 and 3, we obtain
\ba
v_{\rm eff,SW}&=&\left[2-\frac{5}{3}\frac{9/10}{D_\Phi(a_{\rm dec})}\right]D_\Psi(a_{\rm
  dec})\times\frac{3}{2}\cos\theta_e\left[A\left((\chi_e\cos\theta_e-x_c)\frac{\cos^2\theta}{2}+\Delta
    \chi_{\rm
      dec}\frac{\cos^3\theta}{3}\right)\right. \no\\
&+&\left. B\left((\chi_e\cos\theta_e-x_c)^2\frac{\cos^2\theta}{2} +(\chi_e\cos\theta_e-x_c)\Delta\chi_{\rm
      dec}\frac{2\cos^3\theta}{3}+\frac{\Delta\chi_{\rm
        dec}^2 \cos^4\theta}{4}\right)\right]_{\theta_c}^{0}\ .
\ea
\ba
v_{\rm eff,Doppler}&=&{\bf v}({\bf r}_e,a_e)\cdot\hat{n}_e+\frac{3}{2}\cos\theta_e D_v(a_{\rm
  dec}) \\
&\times& \left[\frac{A \cos^3\theta}{3}+\frac{2B(\chi_e\cos\theta_e-x_c)\cos^3\theta}{3}+\frac{B\Delta\chi_{\rm
  dec}\cos^4\theta}{2}\right]_{\theta_c}^{0}\ .\no
\ea
To calculate the ISW effect, we just need to replace $\Psi_i$ in
Eq. \ref{eqn:ISW} by
\ba
\Psi_i&\rightarrow& \frac{3\cos\theta_e}{2} \left[A\left((\chi_e\cos\theta_e-x_c)\frac{\cos^2\theta}{2}+\Delta
    \chi\frac{\cos^2\theta}{3}\right)\right. \no\\
&+&\left. B\left((\chi_e\cos\theta_e-x_c)^2\frac{\cos^2\theta}{2} +(\chi_e\cos\theta_e-x_c)\Delta\chi\frac{2\cos^3\theta}{3}+\frac{\Delta\chi^2 \cos^4\theta}{4}\right)\right]_{\theta_c}^{0}\no
\ .
\ea
Notice that now $\cos\theta_c=(x_c-\chi_e \cos\theta_e)/\Delta \chi
(a)$.  In the limit of no intrinsic CMB dipole, we recover the conventional kSZ effect with $v_{\rm
  eff}={\bf v}({\bf r}_e,a_e)\cdot \hat{n}_e$.

Fig. \ref{fig:zc0}, \ref{fig:zc1}, \ref{fig:zc2}, \ref{fig:zc4},
\ref{fig:vza} \& \ref{fig:vzb} show the values of $v_{\rm eff}$ and
its dependence on $\theta_e$, $z_e$, $z_c$, $A$ and $B$. $z_c$ is the minimum redshift to the edge of bubble collision, with $x_c=\chi(z_c)$. For quick reference, we plot the $x_c$-$z_c$ relation in Fig. \ref{fig:thetaczc}. As expected, $v_{\rm eff}/c$
is in general of the same order as $A$ and $B$. Nevertheless,
numerical results show that the maximum value of $v_{\rm eff}/c$ is often smaller
than $A$ or $B$ by a factor of a few to $10$. This is likely caused
by  incomplete cancellation between the SW, ISW and the Doppler
effect \cite{GZ78,Turner91,Bruni94,Erickcek08}. 

We show $v_{\rm eff}$ as a function of $\theta_e$  for different
values of $z_e=0$, $1$, $2$ \& $4$, and  $z_c=0$ (Fig. \ref{fig:zc0}),
$z_c=1$ (Fig. \ref{fig:zc1}), $z_c=2$ 
(Fig. \ref{fig:zc2}) and $z_c=4$ (Fig. \ref{fig:zc4}). 
$v_{\rm eff}$
shows unique directional dependence on $\theta_e$, which in general
can not be mimicked by conventional sources and
contaminants. Furthermore, such directional dependence relies on
$z_e$, $z_c$, $A$ and $B$  in complicated and different ways. For the purpose of extracting the
bubble collision generated kSZ effect from otherwise overwhelming contaminants and
constraining collision parameters ($A$, $B$ and $x_c\equiv
\chi(z_c)$), this is good news. 

Although in general these features can only be obtained numerically, some of them can be understood analytically. 
\begin{enumerate}
\item For example,
due to the azimuthal symmetry, $v_{\rm eff}(\theta_e=\pi/2)=0$. This holds for all values of $z_e$, $z_c$, $A$, $B$ and $x_c$. Also
for this symmetry, $v_{\rm eff}(\pi-\theta_e)=-v_{\rm eff}(\theta_e)$
when $z_e=0$ and $z_c>0$. When $z_c=0$, this symmetry still holds for the case
B. But for case A, due to the discontinuity in the velocity field at
$z_e=0$, we have $v_{\rm eff}(\pi-\theta_e)=v_{\rm eff}(\theta_e)$
instead.  
\item For case A with $z_c=0$ and $z_e=0$, we have $v_{\rm
  eff}(\theta_e)=v_e(z=0)\cos\theta_e$ when $\theta_e>\pi/2$, and $v_{\rm
  eff}(\theta_e)=-v_e(z=0)\cos\theta_e$ when $\theta_e<\pi/2$. The
value of $v_e(z=0)$ is given in Fig. \ref{fig:v}. 
\item For some
combinations of $z_c$, $z_e$ and $\theta_e$, the LSS of electrons is 
located outside of the region affected by the bubble collision. Therefore we have
$v_{\rm eff}=0$ there. 
\item For some other combinations of $z_c$, $z_e$ and $\theta_e$, the LSS of electrons are located completely inside of the region affected by the bubble collision. Therefore we have
$v_{\rm eff}=0$ for case A, due to the cancellation of the SW, Doppler and ISW effect \cite{GZ78,Turner91,Bruni94,Erickcek08}. 
\item For case A, there is often a discontinuity in
$v_{\rm eff}-\theta_e$. This happens when electrons with varying
$\theta_e$ cross the causal boundary of collision, where the velocity is
discontinuous. 
\end{enumerate}

It is also useful to examine the dependence of $v_{\rm eff}$ on $z_e$ for
fixed $\theta_e$. This, together with $a\bar{n}_e\propto (1+z)^2$ and the evolution of electron number overdensity,  determines the contribution of
electrons at a given redshift  to the kSZ effect. Therefore we plot
the $v_{\rm eff}$-$z_e$ relation in Fig. \ref{fig:vza} \&
\ref{fig:vzb}, for various $\theta_e=0,\pi/4,\pi/2, 3\pi/4$ \&
$\pi$. As expected, $v_{\rm eff}$ along a fixed line of sight is
coherent over Gpc scales. It thus avoids
cancellation of velocity of different signs in the conventional
kinetic SZ effect \cite{Vishniac87}.  Also for this reason, the kSZ-galaxy number density cross correlation does not vanish, in contrast to the vanishing cross correlation between the conventional kSZ and galaxies (e.g. Fig. 1, \cite{Shao11b}).

\subsection{The kSZ auto power spectrum}
The kSZ auto power spectrum induced by a bubble collision is extremely small. For an order of magnitude estimate, we can compute
\begin{equation}
\Delta T/T \sim \tau (v_{\rm eff}/c) \delta_e
\end{equation}
with $\tau\sim 0.01$ and density fluctuation $\delta_e \sim 0.1$ at arcminute scales where the signal will peak. From the previous section, the maximum value of $v_{\rm eff}/c$ is roughly set by the magnitude of $A$ and/or $B$. As we discuss in more detail in Sec.~\ref{eq:cmb_constraints}, existing constraints from the collision signature in the primary CMB are roughly $A < 10^{-4}$ ($B$ has yet to be constrained in the primary CMB). The level of the signature in kSZ power spectrum is therefore $\Delta T^2 \la 1\mu$K$^2$ (for the unrealistic values $A, B \sim 10^{-3}$) to $\Delta T^2\la .01\mu$K$^2$ (at the upper bounds allowed by the primary CMB). This is far smaller than the most recent measurement of the kSZ power spectrum from the South Pole Telescope~\cite{George:2014oba} (SPT) which finds the bandpower $\Delta T^2 = 2.9 \pm 1.3 \mu$K$^2$ at $\ell = 3000$, notwithstanding the primary CMB, thermal SZ, cosmic infrared background, and instrumental noise. 
 
To confirm this rough estimate, we can perform a more accurate computation following the derivation of the dark flow induced kSZ effect outlined in Ref.~\cite{Zhang10d}. Using the Limber approximation, the kSZ power spectrum in the $\hat{n}_e$ direction is given by,
\ba
\Delta T^2_{\rm kSZ}(\ell,\hat{n}_e)=(0.167\mu {\rm k})^2\frac{\pi}{\ell}\int_0^{z_{\rm re}}
\Delta^2_e\left(\frac{\ell}{\chi_e},z_e\right)\left[\frac{v_{\rm eff}(\hat{n}_e,z_e)}{10 {\rm km}/s}\right]^2(1+z_e)^4\chi_ed\chi_e\ .
\ea
$\Delta_e^2(k,z)$ is the power spectrum (variance) of the
electron number overdensity at redshift $z$. In principle it has
contributions both from the conventional primordial fluctuation and
from the bubble collision. Observationally we know that the first
dominates otherwise the concordance $\Lambda$CDM cosmology would not
work so well. Therefore we will approximate $\Delta_e^2$ as that
predicted by the standard $\Lambda$CDM cosmology.\footnote{We have
  numerically verified that this is indeed the case for interesting values of $A$,
  $B\la 10^{-3}$.}  

Fig. \ref{fig:cl} shows the numerical results for the kSZ auto power
spectrum for case B with $B=10^{-3}$ toward the center of the bubble
collision. It peaks at $\ell \sim 10^4$, with peak amplitude $\la 1\muk^2$. It is subdominant to the conventional kSZ effect by a factor of a few at all angular scales. It
is also subdominant to the lensed primary CMB at $\ell\la 4000$.
Fig. \ref{fig:kszA} \& \ref{fig:kszB} show that this sub-dominance
holds for other lines of sight and for both case A and B, with $A,B=10^{-3}$ respectively. The auto power spectrum for other values of $A,B$ scales as $(A,B/10^{-3})^2$.  Since only $A,B\la 10^{-4}$ is allowed by constraints from primary CMB, the kSZ effect induced by bubble collisions cannot be detected in the temperature power spectrum. Nevertheless, we now show that kSZ tomography, namely the kSZ-large scale structure cross correlation, is a promising probe for bubble collisions.

\subsection{The kSZ-large scale structure cross correlation}
Although the kSZ signature of bubble collisions in the temperature power spectrum is hopelessly buried, kSZ tomography can be used to boost the signal by specifically designed weighted cross correlations of CMB temperature with galaxies~\cite{Zhang01,Ho09,Zhang10d,Shao11b}. A schematic is shown in Fig.~\ref{fig:kszfigure}; in a set of redshift bins the kSZ effect caused by bubble collisions is a modulated map of the galaxy distribution onto CMB temperature. This technique has been applied to WMAP-SDSS ~\cite{Li12} and Planck-SDSS (Li, Zhang \& Jing, submitted to ApJ) to search for the dark flow induced kSZ effect. In addition to correlations with the distribution of galaxies, one also expects a tight correlation between the kSZ effect in the CMB and the lensing field (e.g. \cite{Shao11a}) or other tracers of large scale structure such as the 21cm intensity maps. For brevity, we focus in this work on cross correlating with galaxies. 

\bfinew[width=12cm]{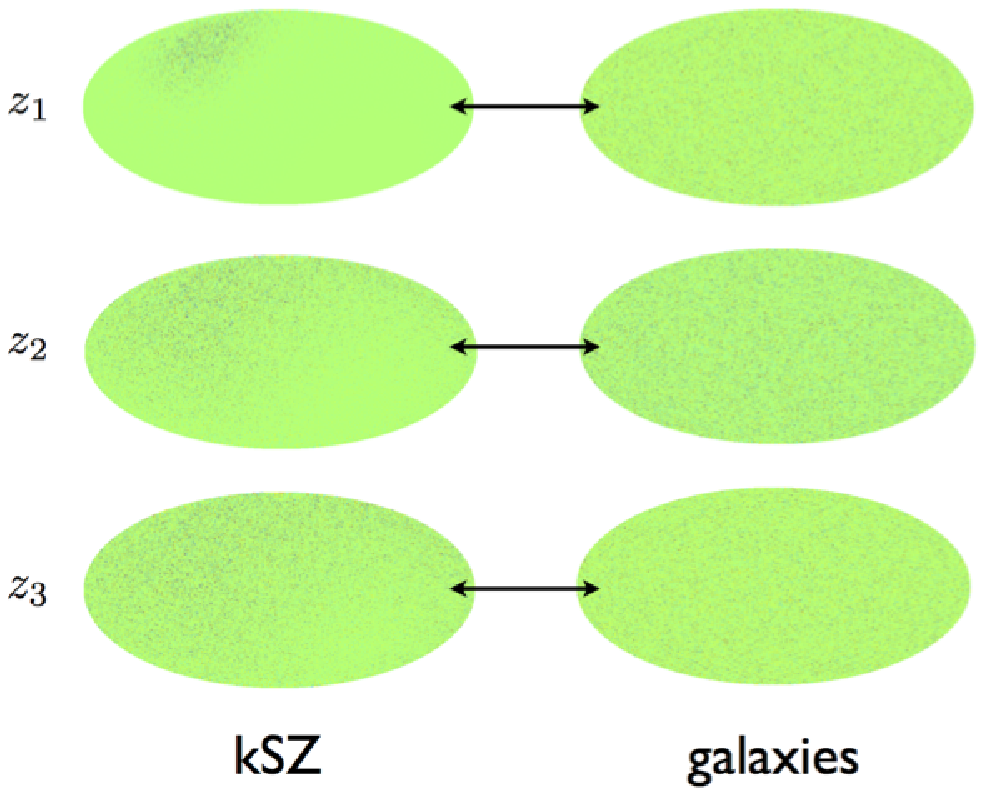}
\caption{A schematic of the correlation between kSZ and the distribution of galaxies. In each redshift bin (vertical), the kSZ effect maps the density of free electrons $\delta_e (\theta,\phi, z_e)$ contained in galaxies (right panel) to CMB temperature (left panel), with a varying amplitude determined by the collision-induced velocity field $v_{\rm eff} (\theta,\phi, z_e)$ (e.g. as shown in Figs.~\ref{fig:zc0}-\ref{fig:zc4}). The measured kSZ effect in CMB alone would be a sum over all redshift bins, however the kSZ-galaxy correlation allows one to reconstruct the velocity field as a function of redshift; this is kSZ tomography.\label{fig:kszfigure}}
\efinew
\bfinew[width=9cm]{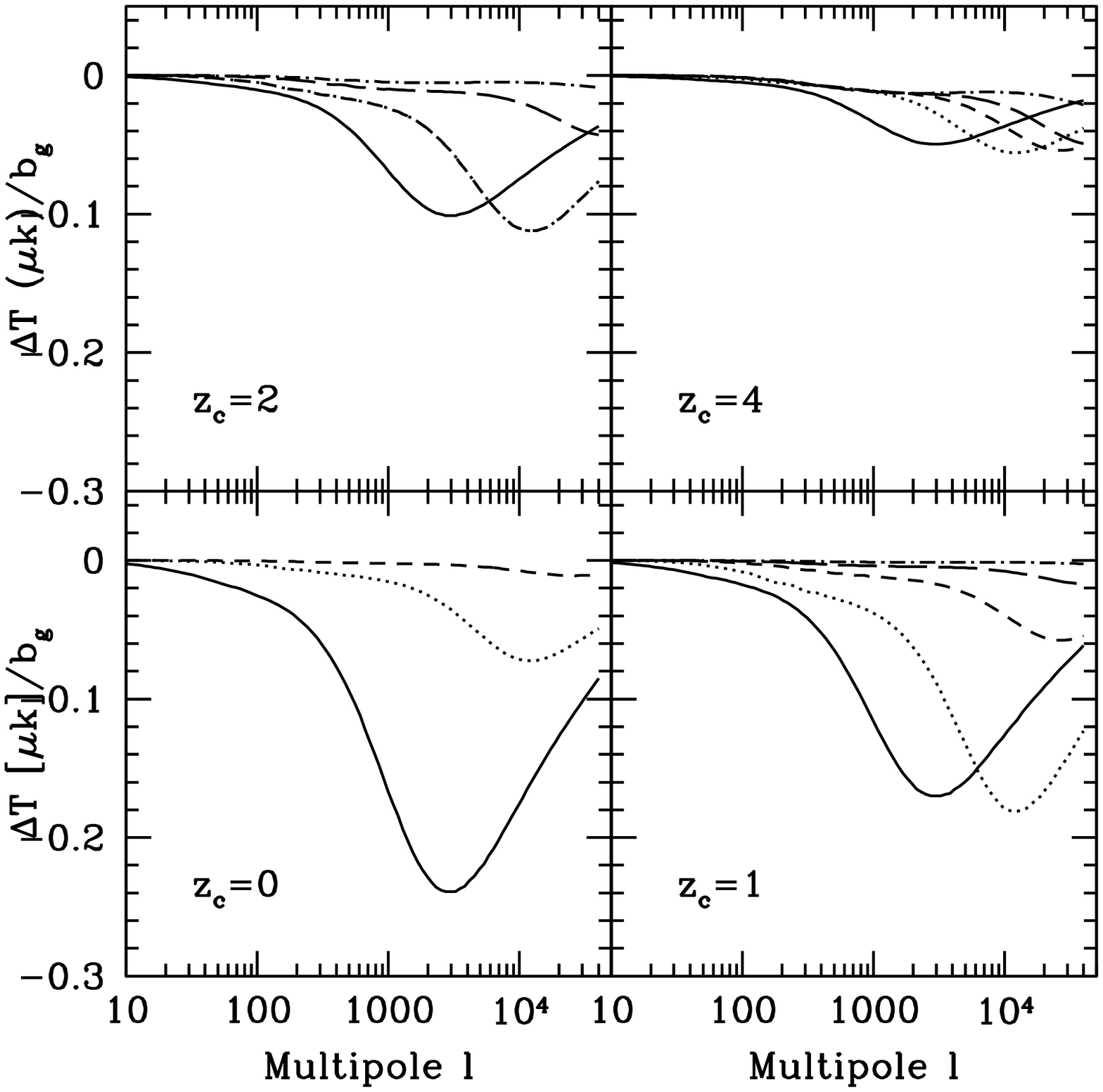}
\caption{The kSZ-galaxy cross power spectra for $B=10^{-3}$ toward the
  center of bubble collision ($\theta_e=0^{\circ}$). Solid,
  dot, short dash, long dash and dot-dash curves have galaxies in the range
  $[0.01,0.3]$, $[0.9,1.1]$, $[1.9,2.1]$, $[2.9,3.1]$, $[3.9,4.1]$,
  respectively. \label{fig:Tgcl}}
\efinew
\bfinew[width=9cm]{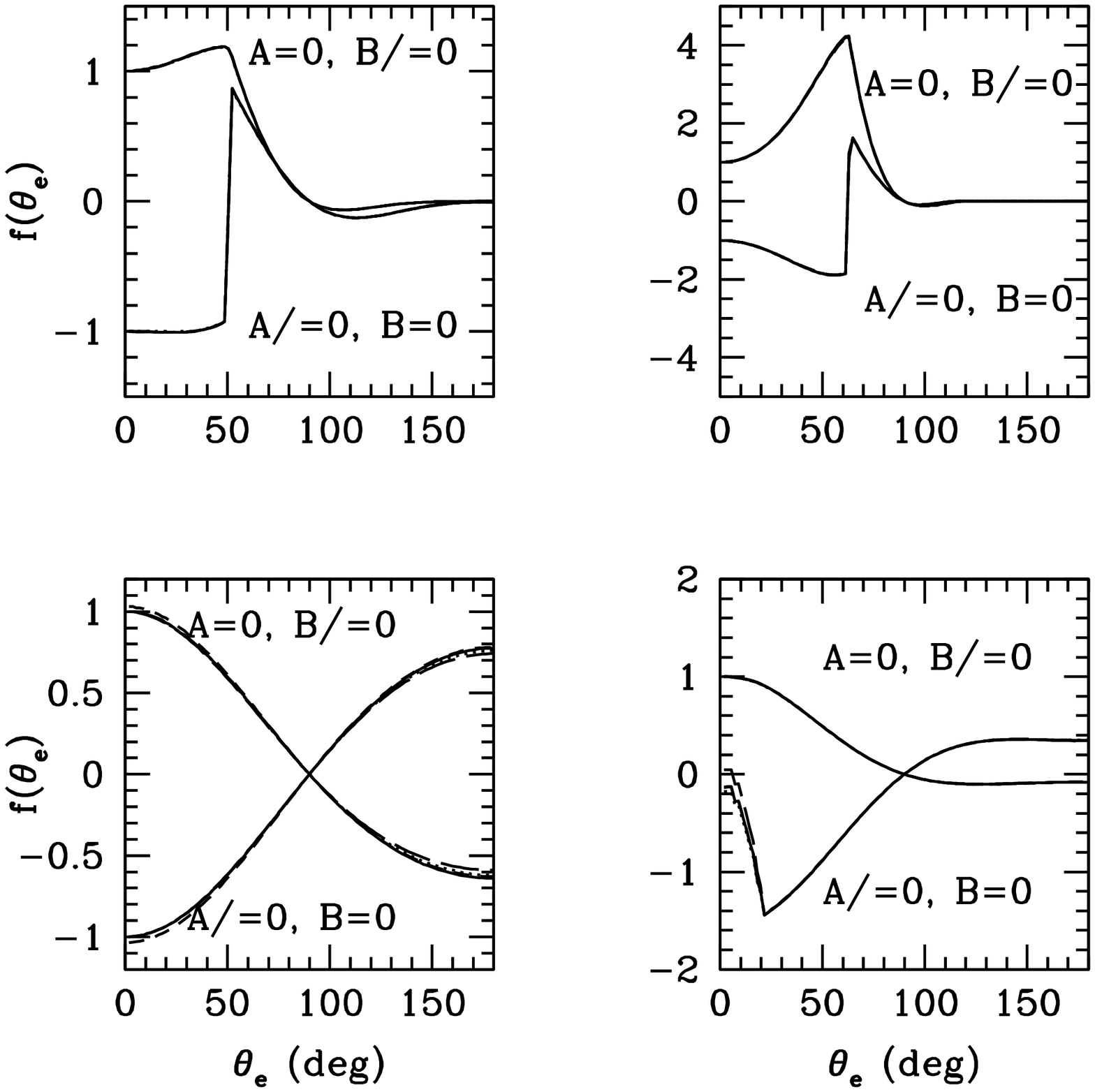}
\caption{The directional dependence of kSZ-galaxy cross correlation
  $f(\theta_e|\ell)$. The panels are for $[0.01,0.3]$ (bottom left), , $[0.9,1.1]$ (bottom
  right), $[1.9,2.1]$ (top left) and  $[3.9,4.1]$ (top right).  In each
  panel, there are four curves for $\ell=1000$ (solid), $3000$ (dot),
  $9000$ (short dash) and the approximation Eq. \ref{eqn:fthetae}
  (long dash), respectively.  All the curves with $A=0$ and $B\neq 0$ are  normalized to unity at $\theta_e=0^{\circ}$. For clarity, curves with $A\neq
  0$ and $B=0$ are normalized to $-1$ at $\theta_e=0^{\circ}$. One
  exception is the redshift bin $[0.9,1.1]$. It has $\Delta
  T(\theta_e=0^{\circ})\simeq 0$. So we normalize with the value at
  $\theta_e=45^{\circ}$ instead. Curves of different $\ell$ overlap over each other and are
  barely distinguishable against each other. This shows that (1) $f(\theta_e|\ell)$ has
  negligible $\ell$ dependence ($f(\theta_e|\ell)\simeq
  f(\theta_e)$) and (2) the approximation Eq. \ref{eqn:fthetae}
  describes $f(\theta_e)$ to excellent accuracy. \label{fig:Tg_ang}}
\efinew

For galaxies in a redshift bin $[z_1,z_2]$, the surface overdensity is $
\delta\Sigma=\int_{z_1}^{z_2}\bar{n}_g\delta_g dz$. 
Here, $\bar{n}_g$ is normalized such that $\int_{z_1}^{z_2}\bar{n}_g
dz=1$. We consider a typical bin size of $\Delta z\equiv z_2-z_1\sim
0.2$. For redshift bins defined by photometric redshift, we shall take
the broadening of redshift width by photo-z error into account. $v_{\rm eff}(\hat{n})$ only varies slowly with sky direction. So to
excellent approximation, over degree scales, it can be treated as
constant. Hence for a degree-sized patch centered on the direction $\hat{n}_P$, the cross correlation of galaxies with kSZ can be well approximated by
\cite{Zhang10d}
\ba
\label{eqn:CTg}
\frac{\ell^2}{2\pi}C_{Tg}(\ell|\hat{n}_P)=-\frac{\pi}{\ell}\int_{z_1}^{z_2}
\frac{\bar{n}_e\sigma_T a}{c} \Delta^2_{eg}(k,z) \bar{n}_g(z)\chi(z) v_{\rm eff}(\hat{n}_P,z)
dz \ .
\ea
Here the electron density-galaxy density power spectrum (variance)
$\Delta^2_{eg}(k,z)$ is evaluated at $k=\ell/\chi(z)$, following the
Limber approximation. We can define a temperature fluctuation $\Delta
T\equiv \ell^2C_{Tg}T_{\rm CMB}/(2\pi)$. We then have
\ba
\Delta T=-0.167 \mu {\rm k} \times \frac{\pi}{\ell}\int_{z_1}^{z_2}\Delta^2_{eg}\left(\frac{\ell}{\chi_e},z_e\right)\left[\frac{v_{\rm eff}(\hat{n}_e,z_e)}{10 {\rm km}/s}\right] (1+z_e)^2\chi_e
\bar{n}_g(z_e)  dz_e\ . 
\ea
Notice that $\bar{n}_g$ is normalized such that $\int_{z_1}^{z_2} \bar{n}_g(z)dz=1$.  The cross
correlation depends on not only the bubble collision parameters, but
also on the galaxy redshift and galaxy bias. For brevity, we only show the result for case B
with $\theta_e=0$ and $B=10^{-3}$, for five redshift bins of $[0.01,0.3]$, $[0.9,1.1]$,
$[1.9,2.1]$, $[2.9,3.1]$ \& $[3.9,4.1]$ (Fig. \ref{fig:Tgcl}). Results for other values of $B$ scale with $B/10^{-3}$. We adopt a simple linear bias for galaxy overdensity, $\delta_g=b_g\delta_m$ and adopt $b_g=1$ unless otherwise specified. $\Delta T$ defined here is not only
proportional to the kSZ signal in the given redshift range, but also
proportional to the density fluctuation in the given redshift
range. Hence, comparing $\Delta T$ over different redshifts can be
misleading. It is better to interpret this as the kSZ signal weighted by
the r.m.s fluctuation of galaxies in the given redshift bin.

\section{Using kSZ tomography to search for bubble collisions}
\label{sec:tomography}
The kSZ-galaxy cross correlation has unique directional dependence,
due to the directional dependence in $v_{\rm eff}$. This characteristic directional dependence is the key to isolating the weak kSZ effect induced by a bubble collision from overwhelming contaminations such as primary CMB and cosmic infrared background. This section focuses on understanding this directional dependence (\S \ref{subsec:direction}), and utilizing the  directional dependence to search for bubble collisions (\S \ref{subsec:weighting}). 

\subsection{The directional dependence of kSZ-galaxy cross
  correlation}
\label{subsec:direction}
To detect a 
characteristic directional dependence of the kSZ-galaxy cross correlation, in observations we can split
the survey sky into patches of a few degrees in  
size.  The measured cross power spectrum depends not only on the multipole
$\ell$, but also the relative direction between the patch center and
the bubble collision center. We denote $C_i(\ell)$ as the 
measured cross power spectrum in the $i$-th patch centered on 
$\hat{n}_i$.  For redshift bins of size $\Delta z\sim 0.2$, $v_{\rm eff}(\hat{n},z)$ does not vary
significantly over the patch and over the redshift range. We can therefore write $C_i(\ell)$ as a spatially modulated power
\be\label{eq:Celldirect}
C_i(\ell)=C_{\rm pivot}(\ell|\hat{n}_{\rm pivot})f(\hat{n}_i) \ .
\ee
where we have chosen to normalize the power against a pivot direction $\hat{n}_{\rm pivot}$, nominally chosen to be the collision centre. Since the velocity as a function of direction does not vary strongly in a degree-size patch, we can neglect the $\ell$ dependence in $f$ that comes with the small angular scale variation of the velocity. From Eq. \ref{eqn:CTg}, we can then approximate $f$ as
\be
\label{eqn:fthetae}
f_i\equiv f(\hat{n}_i)\simeq \frac{\int_{z_1}^{z_2} dz \bar{n}_e a\bar{n}_g\chi
 v_{\rm eff}(\hat{n}_i,z)}{\int_{z_1}^{z_2} dz \bar{n}_e a\bar{n}_g\chi
v_{\rm eff}(\hat{n}_{\rm pivot},z)}\ .
\ee
In Fig. \ref{fig:Tg_ang} we show $f$ computed from the definition Eq.~\ref{eq:Celldirect}; as can be seen in this figure, neglecting the $\ell$ dependence in $f$ is an excellent approximation. Furthermore, it shows that Eq. \ref{eqn:fthetae} excellently describes the directional dependence.  Therefore, given the parameters $A$, $B$, $\hat{n}_c$ and $x_c$, we can predict the variation of cross power from one patch to another, with no need of detailed modelling of electron-galaxy cross correlation. In particular we have $f(\theta_e=\pi/2)=0$, due to the azimuthal symmetry. Later on we will use Eq. \ref{eqn:fthetae} for designing optimal weighting to boost the cross-correlation signal.

\subsection{The optimal weighting for bubble collisions}
\label{subsec:weighting}
For a fixed redshift bin, we combine all available patches on the sky to obtain a linearly weighted cross correlation 
\be
C^W(\ell)=\sum_i C^{\rm obs}_i(\ell)W_i\ .
\ee
Here the measured cross power spectrum is $C^{\rm obs}_i=C_i+N_i$
where $N_i$ is the sum of both statistical and systematic noise. 
The optimal weighting depends on the priors we place on the noise term. We discuss two
limiting cases. One is the optimistic case that all large scale structure related 
components (other than kSZ) in the CMB maps have been removed without bias
by combining multiple frequency bands. In
this limit,  the expectation value of $N_i$ is zero ($\langle
N_i\rangle=0$). The r.m.s error of the weighted correlation is 
\be
\langle N^{W,2}\rangle=\sum_i W_i^2\sigma_i^2\ .
\ee
Here, $\sigma^2\equiv \langle N_i^2\rangle$. The normalization of $W$
is irrelevant. However it is convenient to  fix it in order to find the optimal
weighting. We choose the normalization
\be
C^W(\ell)=C_{\rm pivot}(\ell)\Rightarrow \sum_i W_if_i=1\ .
\ee
The optimal weighting is then 
\be
W_i=\frac{f_i/\sigma_i^2}{\sum f_i^2/\sigma_i^2}\ .
\ee
The second case is more conservative, and perhaps more
realistic. In this case,  there are residual large scale structure related components other than kSZ in the CMB maps. Therefore $\langle N_i\rangle \neq 0$. With a modest prior
that these systematic contaminants are statistically isotropic,  we
need $\sum_i W_i=0$ to eliminate these contaminants in the cross correlation. The optimal
unbiased weighting must minimize 
\be
\sum_i W_i^2\sigma_i^2-\lambda_1\left(\sum_i W_if_i-1\right)-\lambda_2\sum_i W_i
\ee
Here, $\sigma_i^2\equiv \langle N_i^2\rangle-\langle
N_i\rangle^2$. $\lambda_{1,2}$ are the two Lagrangian multipliers for
the constraints $\sum_i W_if_i=1$ and $\sum_i W_i=0$ respectively. The solution is 
\be
W_i=\frac{[f_i-\langle f\rangle_S]/\sigma_i^2}{[\langle
  f^2\rangle_S-\langle f\rangle^2_S]\sum_i 1/\sigma_i^2}\ .
\ee
$\langle f\rangle_S$ is the average over the survey area with the
weighting as follows,
\be
\langle f\rangle_S=\frac{\sum_i f_i/\sigma_i^2}{\sum_i
  1/\sigma_i^2}\ .
\ee
The optimal weighting derived earlier for the case of dark flow \cite{Zhang10d,Li12} is a special case corresponding to identical noise of different patches.  The weighted noise is
\be
\langle N^{W,2}\rangle=\frac{1}{\langle f^2\rangle_S-\langle
  f\rangle_S^2}\times \frac{1}{\sum_i 1/\sigma_i^2}\ .
\ee
The ratio of two weighted noises (case one versus case two) is $1-\langle
f\rangle_S^2/\langle f^2\rangle_S\leq 1$. This inequality confirms our
expectation that case two is more conservative and hence has larger
error. 

To be conservative, we will consider the second case in our forecast. The total signal to noise combining measurements at all $\ell$ in a given redshift bin is
\be
\left[\frac{S}{N}\right]^2_{{\rm tot},z}=\sum_\ell C^2_{\rm pivot}(\ell)\left[\langle f^2\rangle_S-\langle
  f\rangle_S^2\right]\left[\sum_i
1/\sigma_i^2(\ell)\right] \ .
\ee
This assumes that errors in different $\ell$ bins are
uncorrelated. Finally we need to combine measurements for all redshift
bins to obtain the final forecast for signal to noise 
\be
\left[\frac{S}{N}\right]^2_{{\rm tot}}=\sum_z
\left[\frac{S}{N}\right]^2_{{\rm tot},z}\ .
\ee

\bfinew[width=9cm]{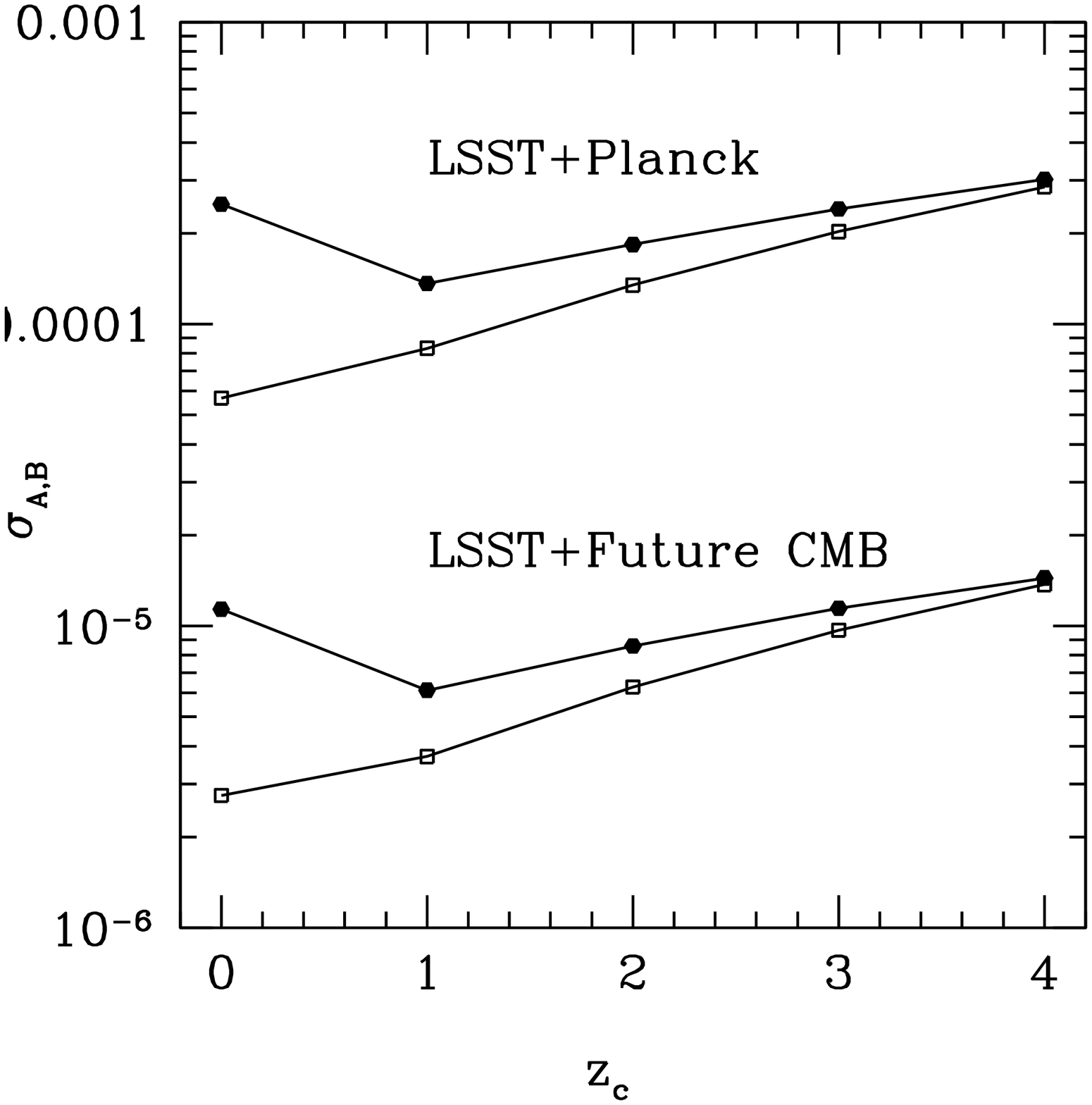}
\caption{The forecasted constrain on bubble collision. For the future
  CMB experiment, we adopt a FWHM=$2^{'}$ and a noise level of
  $1\muk$ per square arcminute. The overlapping sky area is taken as $f_{\rm
    sky}=0.5$. Points with filled circles are for the case of $A\neq
  0$ and $B=0$. Points with open square are for the case of $A=0$ and
  $B\neq 0$. For the angular resolution of Planck, primary CMB
  overwhelms. So Planck in combination with LSST has limited
  capability to detect the kSZ generated by bubble collision. A
  future CMB experiment with better resolution can improve the
  measurement by orders of magnitude. We estimate that a PRISM-like
  experiment has the capability of measure $A$ and $B$ with r.m.s
  error of $10^{-5}$.  Constraints on $B$ are systematically better
  than constraints on $A$ when $z_c\leq 4$, simply reflecting the fact that when
  $A=B$, the case B has on the average larger curvature potential
  inside of our horizon. The two constraints at $z_c=4$ are roughly
  the same, for the reason that they have roughly the same mean
  curvature perturbation inside of our horizon. \label{fig:SN}}
\efinew
We adopt the Gaussian approximation to evaluate $\sigma_i$,
\ba
\label{eqn:sigmai}
\sigma^2_i&=&\frac{C_{Tg,i}^2+(C^{\rm CMB}+C^N_iB^{-2}+C^{\rm kSZ}+C^{\rm For}_i)(C_g+C_g^N)}{2\ell \Delta \ell f_{\rm
      sky,i}}\no\\
&\simeq &\frac{(C^{\rm CMB}+C^N_iB^{-2}+C^{\rm kSZ}+C^{\rm For}_i)(C_g+C_g^N)}{2\ell \Delta \ell f_{\rm
      sky,i}}\ .
\ea
Here, $B=B(\ell)$ is the beam of CMB experiment. $C^N_i$ is the
instrument noise power spectrum at the $i$-th patch of the sky. Due to
variation in the exposure,  $C^N$ varies across the sky. $C^{\rm kSZ}$
is the power spectrum of the total kSZ effect whose major contribution
is the conventional kSZ effect including patchy reionization. $C^{\rm For}$
is the power spectrum of residual foregrounds, which includes all large scale structure related contaminants such as CIB, the relativistic thermal SZ effect and the cosmic radio background.
Since this relies heavily on multi frequency information, whose effectiveness depends
on exposure time, this may also vary with direction. Variation of the gray-body power index and the temperature of CIB over the sky can add more directional variation. Given uncertainties in the modelling of the
conventional kSZ effect and foregrounds, we will adopt a flat spectrum
$\ell^2(C^{\rm kSZ}+C^{\rm For})/(2\pi) (2.73{\rm K})^2=10\muk^2$. $C_g$ is the galaxy
auto power spectrum in the given patch of the sky and the given
redshift bin. $C_g^N$ is the associated (statistical) noise. We only
consider shot noise so that $C_g^N=4\pi f_{\rm
  sky,i}/N_{g,i}$. $f_{\rm sky,i}$ is the fractional sky
coverage. $N_{g,i}$ is the total number of galaxies in this patch of
the sky in the given redshift bin. The last approximation of Eq. \ref{eqn:sigmai} holds since
the kSZ effect induced by bubble 
collision is always subdominant/negligible to sum of other CMB
components/contaminations. 

Given that the bubble collision induced kSZ signal is weak, we require 
both advanced CMB experiments and galaxy surveys in order to be
sensitive to bubble collisions. Fig. \ref{fig:SN} shows the forecasted
constraints on $A$ ($B$), for the combination of Planck and LSST \cite{LSST09}. For LSST, we adopt $\bar{n}_g=30/$arcmin$^2$. The redshift distribution is adopted as $n_g(z)\propto z^2\exp(-z/z_*)$ and $z_*=0.5$. We assume that LSST and Planck overlap over 50\% of the sky. This
combination has the capability of detecting the bubble collision if
max$[A,B]\ga 10^{-4}$. Planck has a limited angular resolution of
$\sim 10^{'}$ ($\ell\ga 2000$), so it misses the majority of the kSZ signal generated
by the bubble collision, which peaks at $\ell\sim 10^4$. To capture the majority of it, the angular  resolution of CMB experiment should be improved to $\sim
2^{'}$. Therefore we consider a PRISM-like CMB experiment \cite{PRISM13} with
FWHM=$2^{'}$ and a low noise level of $1\muk^2$ per
arcmin$^2$. Combined with the LSST survey, the constraint on $A$
and $B$ can be improved by one order of magnitude to $\sigma_{A/B}\ga 10^{-5}$. 

Two major uncertainties in this forecast, besides the bubble collision parameters, are the level of residual foreground contamination and the level of non-Gaussianity of residual foreground near the angular resolution limit. The first can be alleviated by cleaner foreground removal with more frequency bands and lower instrumental noise. A lower residual foreground will make the measurement more noise dominated at small scales, and therefore alleviate uncertainties caused by non-Gaussianities of residual foreground.   

\subsection{Comparison with constraints from the primary CMB}\label{eq:cmb_constraints}

We can compare this sensitivity to what has been achieved using CMB temperature~\cite{Feeney:2012hj,Osborne:2013hea} and what has been forecast for CMB temperature and polarization~\cite{Kleban_Levi_Sigurdson:2011}. These works all analyzed only the linear template, so only provide constraints and projections for the amplitude $A$. CMB studies work with the angular radius of the collision on the CMB sky, given by
\begin{equation}
\cos \theta_c = \frac{x_c}{\chi_{\rm dec}}\ . 
\end{equation}
where $x_c$ is the comoving position of the causal boundary and $\chi_{\rm dec} \simeq 3 H_0^{-1}$ is the comoving distance to the surface of last scattering. To assist in comparison between observables in Fig.~\ref{fig:thetaczc} we show the relation between $x_c$, $z_c$, and $\theta_c$ for the Planck best-fit $\Lambda$ CDM cosmology. 

\bfinew[width=6cm]{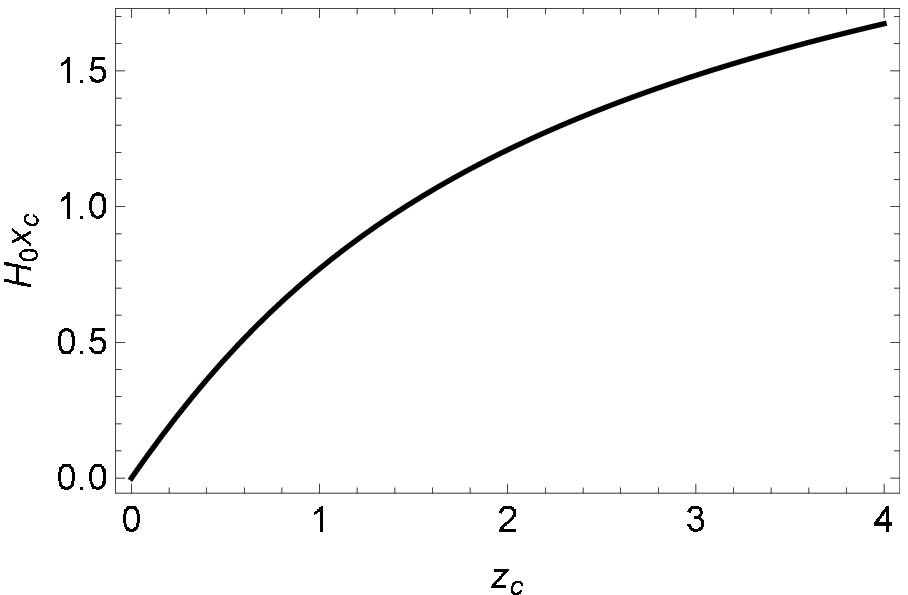}
\includegraphics[width=6cm]{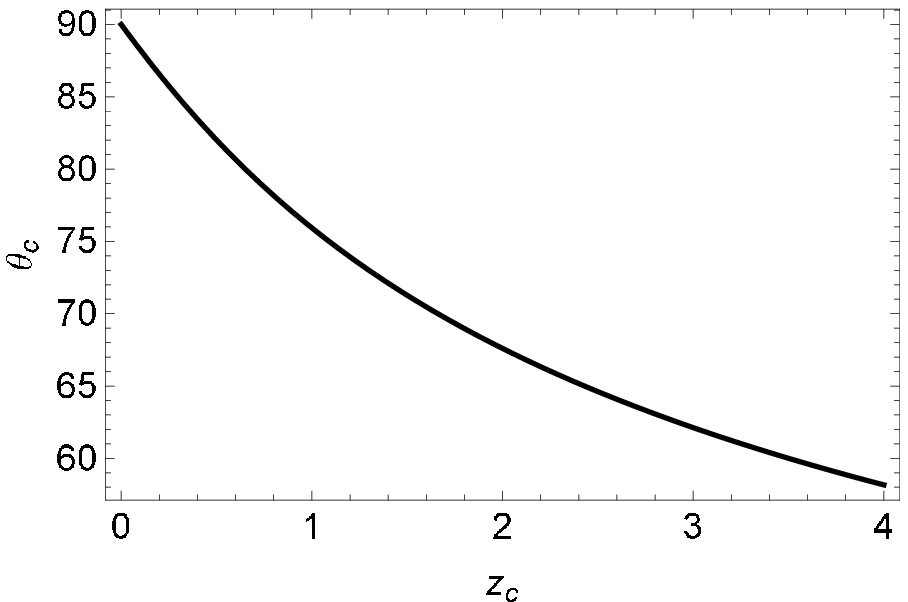}
\caption{The relation between $x_c$ and $z_c$ (left) and $\theta_c$ and $z_c$ (right) in $\Lambda$CDM over the range in redshifts in Fig.~\ref{fig:SN}.  $x_c$ is the x-coordinate of the location of the causal boundary and $z_c$ is the corresponding redshift ($x_c=\chi(z_c)$). $\theta_c$ is the angular radius of the bubble collision on the primary CMB sky ($\cos\theta_c=x_c/\chi_{\rm dec}$). \label{fig:thetaczc}}
\efinew

The sensitivity of the hierarchical Bayesian method of Ref.~\cite{Feeney:2012hj} is given by the 50$\%$ completeness curve for the detection of candidate bubble collisions (Figure 5 of Ref.~\cite{Feeney:2012hj}). The constraint is expressed in terms of $z_0$, the central amplitude of the collision in the CMB temperature. In the Sachs-Wolfe approximation, this is one third of the primordial Newtonian potential $\Psi_i$, and in the notation of Ref.~\cite{Feeney:2012hj} we can identify $A \simeq 3 z_0$. Over the range of $\theta_c$ relevant for Fig.~\ref{fig:SN}, the completeness curve is fairly flat, and we obtain a sensitivity of $|A| \sim 1.5 \times 10^{-4}$. The analysis of candidate features in WMAP7 data failed to provide evidence for bubble collisions, we can interpret this sensitivity as a constraint. 

Ref.~\cite{Osborne:2013hea} performed an analysis of the WMAP7 data using different methods, obtaining the constraint $-2.10 \times 10^{-4} < a H_0^{-1} (\sin \theta_c)^{4/3} < 2.13 \times 10^{-4}$ where $a$ is the amplitude of the linear template in the comoving gauge. On super horizon scales, we can relate this to $A$ in the Newtonian gauge by a factor of $3/5$. We then obtain a constraint on the amplitude A of $|A| < 1.2 (\sin \theta_c)^{-4/3} \times 10^{-4}$. The angular dependence is not important over the range of $\theta_c$ probed by kSZ tomography, so we can express the constraint as $|A| < 1.2 \times 10^{-4}$, in agreement with Ref.~\cite{Feeney:2012hj}. 

Extra information is carried by CMB polarization~\cite{Kleban_Levi_Sigurdson:2011}. Collisions on the angular scales probed by kSZ tomography could be detected at similar signal to noise in CMB polarization as in CMB temperature. The cosmic-variance limited CMB signal to noise using both temperature and polarization will therefore be close to the constraints quoted above.

\subsection{Constraints on the lagrangian underlying eternal inflation}

Auxiliary information is needed to interpret our constraints as constraints on fundamental parameters in the scalar field lagrangian underlying eternal inflation, as outlined in Sec.~\ref{sec:bubblecollisions}. The most important auxiliary observables are the energy density in curvature $\Omega_k^{\rm obs}$ and the scalar to tensor ratio $r^{\rm obs}$. The fundamental limit on observing curvature is $\Omega_k \sim 10^{-5}$, when it becomes indistinguishable from a long-wavelength curvature perturbation. A next-generation CMB polarization experiment can in principle reach sensitivities for the scalar to tensor ratio of $r \sim 10^{-3}$~\cite{Abazajian:2013vfg}, while futuristic space-based gravitational wave detectors may probe down to $r \sim 10^{-6}$~\cite{Friedman:2006zt}. 

The first case to consider is that of collisions between identical bubbles, in which case we can identify $B = 2 \Omega_k^{\rm obs} \left( 1 - \cos \Delta x_{\rm sep} \right)^2 / 15$. In this scenario, kSZ tomography can probe the entire observable range of $\Omega_k^{\rm obs}$. If curvature is detected in the future, kSZ tomography can completely rule out the presence of identical bubble collisions in our observable universe. 

For collisions between non-identical bubbles, the story is not as clear-cut. Generally speaking, we can only constrain  combinations of the fundamental parameters
\begin{equation}
\frac{\delta \phi_0^{\rm coll}}{M_{\rm pl}} \left( 1 - \cos \Delta x_{\rm sep} \right) < \frac{5}{2} \sqrt{ \frac{r_{\rm obs}}{8 \Omega_k^{\rm obs}} } A_{\rm limit}
\end{equation}
\begin{equation}
\sqrt{r_{\rm coll} }   \frac{H_{I}^{\rm coll}}{H_{I}^{\rm obs}} \left( 1 - \cos \Delta x_{\rm sep} \right)^2 < \frac{15}{2} \frac{\sqrt{r_{\rm obs}}}{\Omega_k^{\rm obs}} B_{\rm limit}
\end{equation}
where $A_{\rm limit}$ and $B_{\rm limit}$ are limits on these amplitudes obtained through kSZ tomography (or another method). The first constraint involving $\delta \phi_0^{\rm coll}$ depends on the ratio of $r_{\rm obs}$ and $\Omega_k^{\rm obs}$, both of which could be unobservably small. Smaller values of $r_{\rm obs}$ correspond to small-field models of inflation, which arguably need more tuning to get a large number of $e-$folds, and therefore one might not expect an arbitrarily small $\Omega_k^{\rm obs}$. This is the only theoretical guidance the authors imagine that one might obtain in the absence of a more fundamental understanding of inflation. However, in the optimistic scenario where both curvature and tensors are detected, meaningful limits on fundamental parameters can be set using kSZ tomography.

\section{Discussion and Summary}
\label{sec:discussion}

In this paper we have examined how collisions between bubble universes, a possible relic of eternal inflation, might be observed using the kinetic Sunaev Zel'dovich effect. The kSZ effect we consider is a census of the CMB dipole observed by free electrons on our past light cone, which inhabit nearly every vantage point in our observable universe. Because the kSZ effect is a census, it has the power to probe large-scale inhomogeneities far better than the cosmic variance limited results  associated with observations made from one position. Unfortunately, being a secondary effect, the amplitude of the kSZ signature expected in the CMB is tiny, of order $1 \mu$K. However, a crucial observation is that free electrons trace large scale structure, giving rise to a strong correlation between the kSZ effect and large scale structure in the universe. This redshift-dependent cross-correlation, known as kSZ tomography, can be used as a sensitive probe of ultra-large scale inhomogeneities. 

The effects of a bubble collision are encoded in a planar-symmetric curvature perturbation, and for the class of observers in the vicinity of the causal boundary of the collision, can be described by three constants~\cite{Wainwright:2013lea,Wainwright:2014pta}: the position of the causal boundary (comoving position $x_c$ or equivalently the redshift $z_c$), the amplitude of a linear potential perturbation ($A$), and the amplitude of a quadratic potential perturbation ($B$). The angular dependence of the kSZ effect produced by bubble collisions depends on the parameters $A,B$, and $z_c$. This makes the kSZ effect a powerful discriminator between models, and is essential for isolating the kSZ signature from contaminants. 

The main result of our analysis are summarized in Fig.~\ref{fig:SN}. We forecast that Planck and LSST can put useful constraints $\sigma_{A/B}\sim 10^{-4}$ on the bubble collision parameters $A$ and $B$ over the allowed range of $z_c$. The primary source of this limit come from the angular resolution and instrumental noise of Planck. For LSST and a next-generation CMB experiment such as PRISM, the sensitivity can be improved by roughly an order of magnitude to $\sigma_{A/B}\sim 10^{-5}$. The in-principle limit on sensitivity, in the absence of instrumental noise and assuming perfect foreground subtraction, can be estimated as roughly $\sigma_{A/B}\sim 10^{-10}$.~\footnote{This number is obtained with several approximations and can only be treated as an order of magnitude estimation. (1) We neglect primary CMB. This is indeed valid at $\ell\ga 4000$ where other components dominate. (2) we assume  perfect foreground removal and vanishing instrument noise. (3) we assume sufficiently high galaxy number density such that $C_g^N\ll C_g$. Therefore we can neglect all terms except $C^{\rm kSZ}$ and $C_g$ in Eq.~\ref{eqn:sigmai}. (4) we assume a perfect cross correlation between the kSZ effect in a given redshift bin and the galaxy overdensity in that redshift bin, namely the cross correlation coefficient $r=1$. These approximations allow us to sum over all $\ell$ and $z$ bins and obtain $(S/N)^2_{\rm tot}\sim \ell^2_{\rm max}f_{\rm sky} C^{\rm kSZ}_{\rm bubble}/C^{\rm kSZ}$. We first consider case B. Using $\ell^2C^{\rm kSZ} _{\rm bubble}/(2\pi) \sim B \times 10^2 \mu$K  (Fig. \ref{fig:cl}) and  $\ell^2C^{\rm kSZ}/(2\pi) \sim 1 \mu$K for all $\ell$,  we obtain  $(S/N)^2_{\rm tot}\sim 10^{10} B (\ell_{\rm max}/10^4)^2f_{\rm sky}$, valid for $|B|\ll 1$. For $\ell_{\rm max}=10^4$, we obtain a sensitivity of $\sigma_B\sim 10^{-10}$. Constraint on $A$ is of the same order of magnitude. The reason is that the kSZ power spectrum for case A is comparable to case B when $A=B$ (Fig. \ref{fig:kszA} versus Fig. \ref{fig:kszB}),  and it scales linearly with $A$. Nevertheless, constraint on $A$ can be a factor of $\sim 2$ weaker than $B$ due to a factor of $\sim 2$ weaker signal (Fig. \ref{fig:kszA} versus Fig. \ref{fig:kszB}).  } Nearly cosmic variance limited constraints exist for $A$ from the primary CMB~\cite{Feeney_etal:2010dd,Feeney_etal:2010jj,Feeney:2012hj,McEwen:2012uk}, yielding $A \lesssim 10^{-4}$. Therefore, kSZ can in principle  beat the cosmic-variance limited constraints from the primary CMB by nearly 6 orders of magnitude. To put this in perspective, using e.g. $A \sim \sqrt{\Omega_k^{\rm obs}} \sim e^{-N_{\rm extra}}$, where $N_{\rm extra}$ is the number of extra $e$-folds beyond the number necessary to solve the horizon problem, it would be possible at this sensitivity to probe the initial conditions for inflation up to 23 $e-$folds beyond what is necessary to solve the horizon problem.

To connect the constraints on $A$ and $B$ to constraints on the fundamental parameters in the scalar field lagrangian underlying eternal inflation, ancillary data is needed. In particular, a detection of curvature and primordial tensor modes would allow for a direct constraint on the width of the potential barrier describing the collision bubble and the relative energy scale of inflation inside the collision and observation bubbles. In the most constraining scenario ($\Omega_k^{\rm obs} \sim 10^{-2}$, $r_{\rm obs} \sim 10^{-6}$), the in-principle sensitivity of the kSZ effect would allow for constraints on the properties of the collision bubble down to scales of order $10$-$10^3$ TeV ($\delta \phi_0^{\rm coll} < 10^3$ TeV, $\sqrt{r_{\rm coll}} H_{I}^{\rm coll} < H_I^{\rm obs} 10^{-10} < 10$ TeV). These scales are potentially relevant for supersymmetric or grand unified theories which give rise to eternal inflation.

Bubble collisions in eternal inflation are a predictive theory of inhomogeneous initial conditions for our Universe. In this respect, they provide a testbed for addressing more generally what we might learn about the observability of inflationary initial conditions by providing a benchmark model. Clearly, kSZ tomography has enormous potential to probe the initial conditions for inflation.

{\bf Acknowledgment}.---PJZ was supported by the National
Science Foundation of China 
(Grant No. 11025316, 11320101002, 11433001), National
Basic Research Program of China (973 Program 2015CB857001), the
Strategic Priority Research Program "The Emergence of Cosmological
Structures" of the Chinese Academy of Sciences (Grant
No. XDB09000000), the 
NAOC-Templeton beyond the horizon program, and the key laboratory grant from the Office of Science and Technology, Shanghai Municipal Government (No. 11DZ2260700). Research at Perimeter Institute is supported by the Government of Canada through Industry Canada and by the Province of Ontario through the Ministry of Research and Innovation. 
MCJ is supported by the National Science and Engineering Research Council through a Discovery grant. PJZ and MCJ thank the support of
National Science  Foundation Grant No. PHYS-1066293, the Simons
foundation, and the hospitality of the Aspen Center for Physics where
this work was initiated.
 
\appendix
\section{Linear Evolution of gravitational potential and velocity}\label{sec:appendix}

\bfinew[width=9cm]{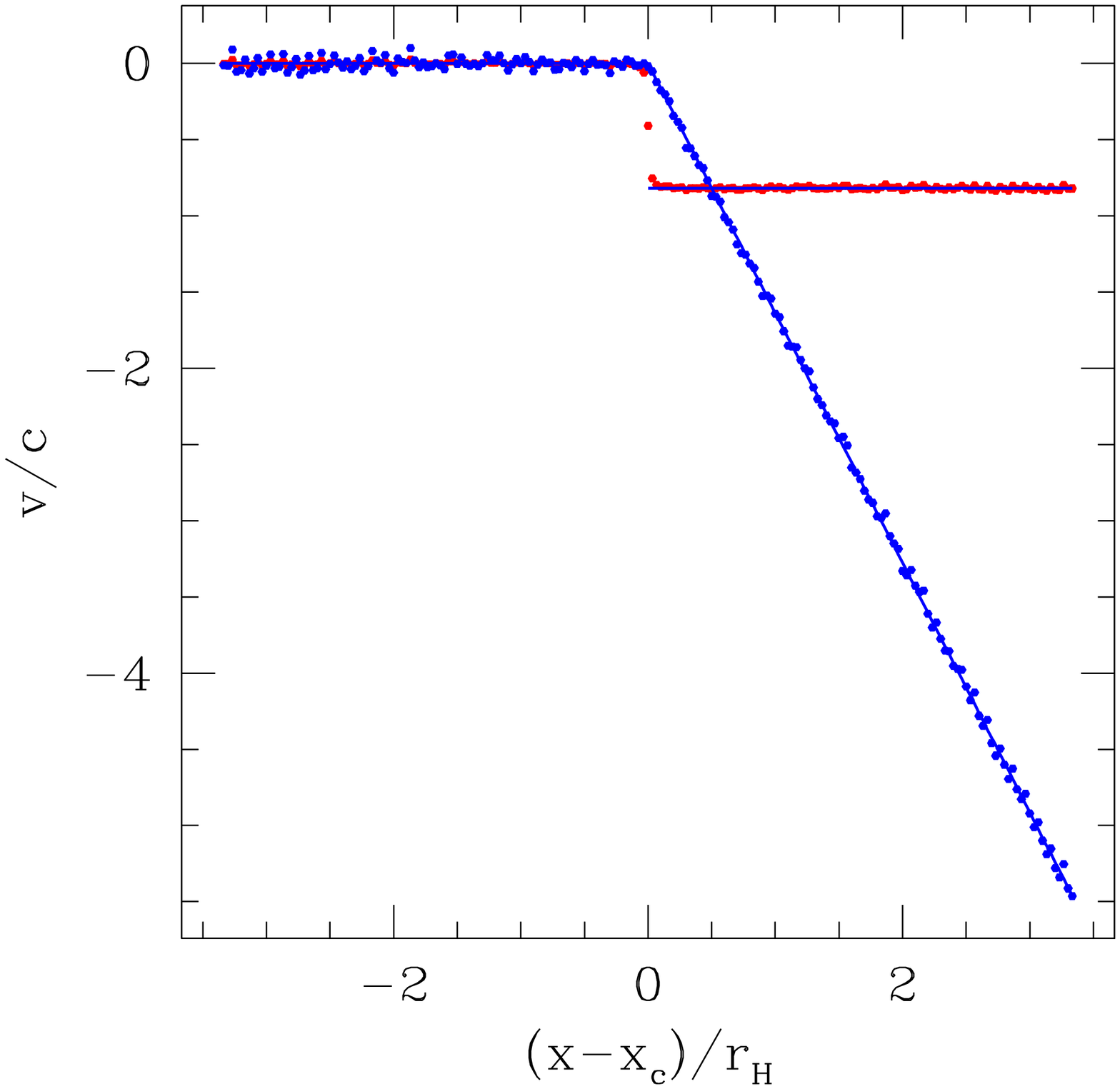}
\caption{The velocity field at $z=0$, for case A and case B. The data points are the
  numerically evaluated velocity field and the solid lines are the
  analytical prediction assuming superhorizon evolution. The two agree
  excellently. Hence  we will adopt the superhorizon
  evolution throughout the paper, which allow us to address many calculations analytically. \label{fig:v}}
\efinew

The bubble collision generated perturbation is a mixture of both
superhorizon and subhorizon modes. Therefore the exact evolution of
$\Psi$ and $v$ has to be solved numerically. We adopt the conformal
Newtonian gauge and solve the following equations for radiation and
cold dark matter in a flat universe \cite{Dodelson03},
\ba
\frac{d\Theta_{r,0}}{dx}&=&-\frac{k}{aH}\Theta_{r,1}-\frac{d\Psi}{dx}\
,\\
\frac{d\Theta_{r,1}}{dx}&=&\frac{k}{3aH}\Theta_{r,0}-\frac{k}{3aH}\Psi\
,\no\\
\frac{d\delta}{dx}        &=& -i\frac{k}{aH}v-3\frac{d\Psi}{dx}\ ,\no \\
\frac{dv}{dx}               &=&-v+i\frac{k}{aH}\Psi\ ,\no \\
\frac{d\Psi}{dx}&=&\frac{1}{2}\frac{H_0^2}{H^2}(\Omega_ma^{-3}\delta+4\Omega_ra^{-4}\Theta_{r,0})-\Psi-\frac{k^2}{3a^2H^2}\Psi\ .\no
\ea
We have changed the argument from the conformal time $\tau$ to $x=\ln
a$.  We numerically solve the above five equations for the variables $(\Theta_{r,0}, \Theta_{r,1}, \delta, -iv, \Phi)$ using a modified version of the CMBFAST package \cite{CMBFAST}.  The adiabatic initial condition is
\ba
\Theta_{r,0}=\frac{1}{2}\Psi_i, \delta=\frac{3}{2}\Psi_i,
\Theta_{r,1}=0, v=0\ .
\ea

We find that, although there are subhorizon modes in the initial
perturbation, the overall evolution of $\Phi$ and $v$ are excellently
described by the superhorizon evolution (Fig. \ref{fig:v}).  For superhorizon modes, the
evolution in potential at early 
time where the cosmological constant is negligible  is \cite{Dodelson03,Erickcek08}
\be
\Psi_{\rm SH}(a)=\Psi_{\rm
  SH,i}\frac{16\sqrt{1+y}+9y^3+2y^2-8y-16}{10y^3}\ .
\ee 
Here, $y\equiv a/a_{\rm eq}$ and $a_{\rm eq}=\Omega_R/\Omega_m$ is the scale factor at
the epoch of  the radiation-matter equality. When $y\gg 1$, $\Psi_{\rm
  SH}\rightarrow \Psi_{\rm SH,i} 9/10$.  On the other hand, the
evolution of gravitational potential at late time where radiation is
negligible is
\be
\Psi(a)\propto \frac{5}{2}\Omega_m \frac{E(a)}{a}\int_0^a
\frac{da^{'}}{E^3(a)a^3}\ .
\ee
Here
$E(a)=\sqrt{\Omega_ma^{-3}+\Omega_\Lambda}$
is the normalized Hubble parameter and  by setting $\Omega_R=0$. This
result holds for both the subhorizon and superhorizon modes
\cite{Zhang11a}. The
integral above can be excellently fitted by \cite{Carroll92}
\be
g(a)\equiv
\frac{5\Omega_m(a)/2}{\Omega_m^{4/7}(a)-\Omega_\Lambda(a)+(1+\frac{\Omega_m(a)}{2})(1+\frac{\Omega_\Lambda(a)}{70})}\ .
\ee
Hence the evolution of a superhorizon mode over all cosmic epoch can be well
approximated by
\ba
\label{eqn:PhiSH}
D_\Psi(a)\equiv \frac{\Psi_{\rm SH}(a)}{\Psi_{\rm
  SH,i}}&=&\frac{16\sqrt{1+y}+9y^3+2y^2-8y-16}{10y^3} \left[ \frac{5}{2}\Omega_m \frac{E(a)}{a}\int_0^a
\frac{da^{'}}{E^3(a)a^3} \right] \\ 
&\simeq& \frac{16\sqrt{1+y}+9y^3+2y^2-8y-16}{10y^3} \times g(a)\ . \no
\ea
On the other hand, we have \cite{Erickcek08}
\ba
\label{eqn:vSH}
{\bf v}=-\frac{2a^2c^2 H(a)}{H^2_0\Omega_m} \frac{y}{4+3y}
\left[\nabla \Psi+\frac{d\nabla\Psi}{d\ln a}\right]= -\frac{2a^2c^2H(a)}{H^2_0\Omega_m}  \frac{y}{4+3y}
\left[D_\Psi+\frac{dD_\Psi}{d\ln a}\right]\nabla \Psi_{i}\ . 
\ea
We define the velocity growth rate function as
\be
\label{eqn:Dv}
D_v(a)\equiv \frac{2a^2H(a)}{H^2_0\Omega_m} \frac{y}{4+3y}
\left[D_\Psi+\frac{dD_\Psi}{d\ln a}\right]\ .
\ee
$D_v$ and $D_\Psi$ are related by
\be
D_v=\frac{1}{a}\int_0^a \frac{D_\Psi(a)}{aH}da\ .
\ee
We find that Eq. \ref{eqn:PhiSH} and \ref{eqn:vSH} describe the
evolution of the given gravitational potential (Eq. \ref{eqn:iPhi})
and the induced velocity excellently (Fig. \ref{fig:v}). Numerically we find that subhorizon modes slightly smooth the potential and velocity near the causal boundary. But these modes are negligible for the purpose of this paper. Therefore we
will adopt the above  analytical results for superhorizon modes to follow the evolution of bubble
collision generated perturbation. 

\bibliography{ksz_bubbles}
\end{document}